\documentclass[aps,prl,showpacs,longbibliography]{revtex4-2}
\usepackage{amsfonts,amsmath,bbm,wasysym}
\usepackage{amsfonts,dsfont}
\usepackage{multirow}
\usepackage{amssymb}
\usepackage[english]{babel}
\usepackage{float}
\usepackage{lipsum}
\usepackage{graphicx}
\usepackage{epsfig}
\usepackage{relsize}
\usepackage{braket}
\usepackage[colorlinks=true,citecolor=red,urlcolor=blue]{hyperref}
\usepackage{soul}
 
\usepackage{ulem}
 \usepackage{bm}

\begin{document}
\title{Realization of staircase topological Anderson phase transitions}
\author{Marwa Manna\"i$^{1}$}
\author{Yaoyao Shu$^{2}$}
\author{Sonia Haddad$^{3}$}
\email{sonia.haddad@fst.utm.tn}
\author{Mina Ren$^{4}$}
\author{Hong Chen$^{2}$}
\author{Yong Sun$^{2}$}
\email{yongsun@tongji.edu.cn}
\author{Hisham Sati$^{1,5,6}$}
\email{ha56@nyu.edu}
\affiliation{\mbox{1} Center for Quantum and Topological Systems, NYUAD Research Institute, New York University Abu Dhabi, UAE }
\affiliation{\mbox{2} MOE Key Laboratory of Advanced Micro-Structured Materials, School of Physics Science and Engineering, Tongji University, Shanghai 200092, China}
\affiliation{\mbox{3} Laboratoire de Physique de la Mati\`ere Condens\'ee, Facult\'e des Sciences de Tunis, Universit\'e Tunis El Manar, Campus Universitaire 1060 Tunis, Tunisia}
\affiliation{\mbox{4} Department of Physics, Shanghai Normal University, Shanghai 200234, China}
\affiliation{\mbox{5} Mathematics, Division of Science, New York University Abu Dhabi (NYUAD), Abu Dhabi, UAE}
\affiliation{\mbox{6} The Courant Institute for Mathematical Sciences, NYU, NY 10012, USA}
\begin{abstract}
    One-dimensional topological Anderson insulators provide a paradigm for disorder-induced topological phases in which the underlying system turns from a trivial to a topological phase. It is widely recognized that the latter vanishes at large disorder amplitude. Here, and contrary to the general belief, we provide evidence for a successive disorder-driven topological transitions in a single-wall nanotube, culminating in a topological Anderson phase that remains unexpectedly robust at strong disorder. This phenomenon is confirmed by analysis of the corresponding topological invariant, which increases stepwise as disorder increases, giving evidence for the emergence of edge states. We experimentally implement these topological Anderson staircase phase transitions in a one-dimensional topolectrical circuit, where the persistence of edge states is revealed by node-voltage measurements. The robustness of the edge states is corroborated by numerical calculations of their localization properties. Our work opens the road to \textit{topological disordertronics}, where topological phases can be tuned by disorder.
\end{abstract}
\maketitle
\section{Introduction}
Disorder is commonly known to thwart quantum transport and to induce Anderson localization \cite{Andersonpaper,Andersonpaper2}. This is not the case of topological phases of matter hosting edge states, which are symmetry protected \cite{Eduardo2025,DisorderHot1,Eduardo2024,Mannai2023,DisorderHot2}.
The robustness of these boundary states against disorder-induced backscattering is the hallmark property of topological insulators. However, at sufficiently strong disorder, the edge states collapse and the system turns into a topologically trivial phase.

\smallskip
A counterintuitive disorder-driven topological phase transition has been proposed to take place in the so-called topological Anderson insulators (TAIs) \cite{TAI1,Beenakker,gappedTAI1,gappedTAI2,gappedTAI3,gappedTAI4,gappedTAI5} where extended edge states are created by disorder. The latter drives the system into a phase transition from a topologically trivial to a non-trivial phase. In contrast to conventional topological insulators, where protection is usually phrased in terms of a spectral bulk gap, TAIs are understood in terms of the mobility gap, where the edge modes remain protected as long as there are no extended bulk states at the same energy.\\
Within this framework, the standard TAI exhibits a reentrant scenario: Starting from a trivial insulator, increasing disorder produces a gapped TAI, then a gapless TAI where localized bulk states coexist with edge modes within the mobility gap \cite{UngappedTAI,ungappedTAI1}, and finally a trivial Anderson insulator in which all states are localized and the edge protection is lost.\\
TAIs have been realized in different backgrounds, including one-dimensional (1D) cold atom lattices  \cite{observation}, photonics \cite{Photonic1,Photonic2,Photonic3,Photonic4,phononic}, acoustics resonators \cite{acoustic}. Higher-order TAIs have been implemented in electric circuits \cite{Zhang2021}, where corner states induced by disorder have been observed. \\
Recently, a model of successive transitions in higher-order TAIs has been proposed \cite{Successive} where the number of the boundary-localized corner states is expected to show a stepped increase. However, this cascade of topological phase transitions is found to collapse at large disorder amplitude, driving the system into a reentrant trivial phase \cite{Successive}.\\
The natural question which arises here is wether multiple phase transitions can occur in TAIs with non-reetrant robust phase surviving in the strong disorder regime. Realizing such a cascade of topological phase transitions will open the gate to topological \textit{disordertronics}: a disorder-tuned topological quantum transport.

\smallskip
In this paper, we theoretically and experimentally address this question in a chiral disordered 1D system where chiral symmetry is preserved. We propose a model describing a single-wall carbon nanotube (SWNT) \cite{topologyCNT1,topologyCNT2,topologyCNT3,topologyCNT4,topologyCNT5} where disorder is introduced only in the long-range intercell couplings, while the intracell bond remain unaffected. We then realize our model in a suitably programmable topolectrical circuit, in which disordered long-range hoppings are implemented by tunable coupling capacitors. We provide evidence of a disorder-induced cascade of topological Anderson transitions where the corresponding topological invariant exhibits a staircase behavior and, eventually saturates at the largest value of disorder we access, without turning the system to a trivial Anderson insulator. The robustness of the unconventional TAI phase, emerging in the strong disorder regime, is further supported by a detailed analysis of several localization parameters, including the zero- energy localization length, the ratio between the geometric- mean and arithmetic-mean density of states (DOS), as well as the inverse participation ratio and the Shannon entropy.\\
Experimentally, we construct a node-voltage measurement for which the LTspice simulations reveal the signature of non-reentrant TAI staircase phase transitions, including successive edge-localized resonances near the zero-energy frequency which remain robust even at strong disorder. In stark contrast with conventional TAI, we do not observe the reentrance of the trivial phase at large disorder.\\
It is worth to note that our work does not only provide an implementation of disorder-driven Anderson topological phase transitions in simulated 1D systems but it also opens the way to realize these transitions in SWNT. Recent studies have shown that some semiconducting SWNTs can host topologically protected edge modes \cite{topologyCNT1,topologyCNT2,topologyCNT3,topologyCNT4,topologyCNT5}. These nontrivial phases belong to class BDI in the tenfold way topological classification, characterized by a $\mathbb{Z}$- invariant \cite{altland1,altland2} indicating the number of edge states. External magnetic fields, have been widely employed to manipulate the topological edge states. However, this occurs at unreachably strong magnetic field \cite{topologyCNT3}. Here, we show that SWNTs are excellent candidates to observe disorder-induced multiple topological phase transitions which can survive at relatively strong disorder amplitude.
\section{Results}
\subsection{Model Hamiltonian} 
SWNTs are hollow cylindrical tubes created by wrapping a monolayer of graphene along the chiral vector $\mathbf{C}= n \mathbf{a_1} + m \mathbf{a_2}$, where $n$ and $m$ are integers and $\mathbf{a_{1,2}}$ are the primitive vectors of the hexagonal honeycomb lattice. Within the standard helical-angular construction for $(n,m)$-SWNTs~\cite{topologyCNT2,topologyCNT3}, the graphene sheet is unrolled and mapped onto $d=\mathrm{gcd}(n,m)$ spiral chains along a helical vector $\mathbf{H}$, so that the circumferential angular momentum $\mu \in \left\{0, ..., d-1 \right\}$ is a good quantum number. For each fixed $\mu$, the system reduces to an effective 1D chiral chain with two atoms per unit cell and $\mu$-dependent long-range hoppings~\cite{topologyCNT2,topologyCNT3}. In the clean limit, these chains belong to symmetry class BDI of the tenfold way topological classification \cite{altland1,altland2} and are characterized by an integer winding number $\omega_\mu$, whose values are determined by $(n, m, \mu)$ (see Supplementary Note 1). Here, we focus on semiconducting nanotubes with $\mathrm{mod} (2n +m,3) \neq 0 $ \cite{topologyCNT3,SWNTsbasic} where the reduced 1D model is a  bipartite chiral-chain. It is worth to note that metallic SWNTs can be described by an effective 1D model with different hopping processes between atoms on the same sublattice, which breaks the simple bipartite picture~\cite{topologyCNT3}. 

\smallskip
The tight-binding Hamiltonian of the reduced 1D semiconducting SWNT with disorder incorporated into the intercell hopping terms reads 
\begin{equation}\label{SWNT-offDiagonal}
    H_\mu= \sum_{l=1}^N J_1 c^{\mu \dagger}_{A,l} c^{\mu}_{B,l} 
    + J_{2, \mu} c^{\mu \dagger}_{A,l} c^{\mu}_{B,l+\Delta'_2}
    +  J_{3, \mu} c^{\mu \dagger}_{A,l} c^{\mu}_{B,l+\Delta'_3}+ \mathrm{H.c.}
\end{equation}
in which $c^{\mu \dagger}_{A(B),l}$ and $ c^{\mu}_{A(B),l}$ are the creation and annihilation operators on sublattice $A (B)$ in the $l$-th unit cell of the $\mu$-th angular momentum chain. We assume that disorder does not affect the nearest-neighbour intracell hopping amplitude denoted $J_1=t$. We discuss the reasons behind such assumption in the following subsection.

The second and third terms describe long-range couplings between the $l$-th and $(l+\Delta_2')$-th unit cells, and between the $l$-th and $(l+\Delta_3')$-th unit cells, respectively, with $\Delta_2'= n/d,\ \Delta_2^{\prime \prime}=-p,\ \Delta_3'=-m/d,$ and $\Delta_3^{\prime \prime}=q$. Here $p$ and $q$ are two integers satisfying the relation $mp-nq=d$. The disordered intercell hoppings are taken as
\begin{equation}
    J_{2, \mu}= t e^{\frac{i 2 \pi \mu \Delta^{\prime \prime}_2}{d}} 
    +W^{\prime} \epsilon^{\prime}_l, \quad J_{3, \mu}= t e^{\frac{i 2 \pi \mu \Delta^{\prime \prime}_3}{d}}+W \epsilon_l.
    \label{J2J3}
\end{equation}
$\epsilon_l$ and $\epsilon_l^{\prime}$ are randomly generated numbers uniformly distributed in the interval $[-0.5, 0.5]$ without correlations, and $W$ and $W^{\prime}$ are the dimensionless disorder strengths. Throughout, we set $W^{\prime} = \beta W$ to control the relative strength of disorder on the two long-range bonds.

\smallskip 
As a concrete example, we hereafter consider the case of $(8,6)$ nanotube. For this chirality, the system can be mapped into two chains, each one has a given angular-momentum index $\mu=0$ and $\mu=1$. In the clean limit, these chains have long-range couplings $(J_{2,\mu}, J_{3,\mu})= (t,t)$ for $\mu=0$ and $(J_{2,\mu}, J_{3,\mu})= (-t,-t)$ for $\mu=1$, corresponding to winding numbers $\omega_0=0$ and $\omega_1=1$, respectively (see Supplementary Note 1). Since we are interested in disorder-induced topology, we focus on the $\mu=0$ chain, which is topologically trivial in the clean limit. Note that one can consider other sets of  semiconducting chiralities $(n,m)$ to study the trivial-topological transitions, as discussed in the Supplementary Note 2.
\subsection{Multiple topological phase transitions} 
The chiral symmetry $\hat\Gamma=I_N\otimes \sigma_z$ remains preserved in the presence of the off-diagonal disorder, with the $N$-rank identity matrix $I_N$ and $\sigma_z$ is the pseudo-spin Pauli matrix describing the honeycomb sublattices. Therefore, protected nontrivial topological phases are still possible. To uncover the disorder-driven topological transitions in a system with broken translational invariance, a real-space winding number can be evaluated using Prodan’s formula \cite{Prodan}. For a chiral symmetric Hamiltonian, the winding number is given by~\cite{Prodan}
\begin{equation}
    \omega_\mu= \frac{1}{L^{\prime}}\text{Tr}^{\prime} 
    \left(\hat\Gamma \hat P[\hat P,\hat X]\right),
\end{equation}
with 
$\hat P=\sum_{j=1}^N \big(\ket{\phi_j} \bra{\phi_j}-\hat\Gamma\ket{\phi_j} \bra{\phi_j}\hat\Gamma^{-1}\big)$ being the flattened version of $H_\mu$, while $\ket{\phi_j}$ is the $j$-th eigenstate of $H_\mu$, satisfying $H_\mu\ket{\phi_j}=E_j\ket{\phi_j}$. Here $X=\sum_j j \big(c^{\mu \dagger}_{A,j} c^{\mu}_{A,j}+ c^{\mu \dagger}_{B,j} c^{\mu}_{B,j}\big)$ is the coordinate operator, and $\text{Tr}^{\prime}$ is the trace over the $L^{\prime}$ unit cells in the middle of the chain to suppress the boundary effects. In addition, we select a sufficiently large system size to ensure self-averaging and avoid finite-size effects. A systematic finite-size scaling analysis of the winding number is provided in Supplementary Note 3.
\begin{figure}[hpbt] 
$\begin{array} {ccc}
\includegraphics[width=0.33\columnwidth]{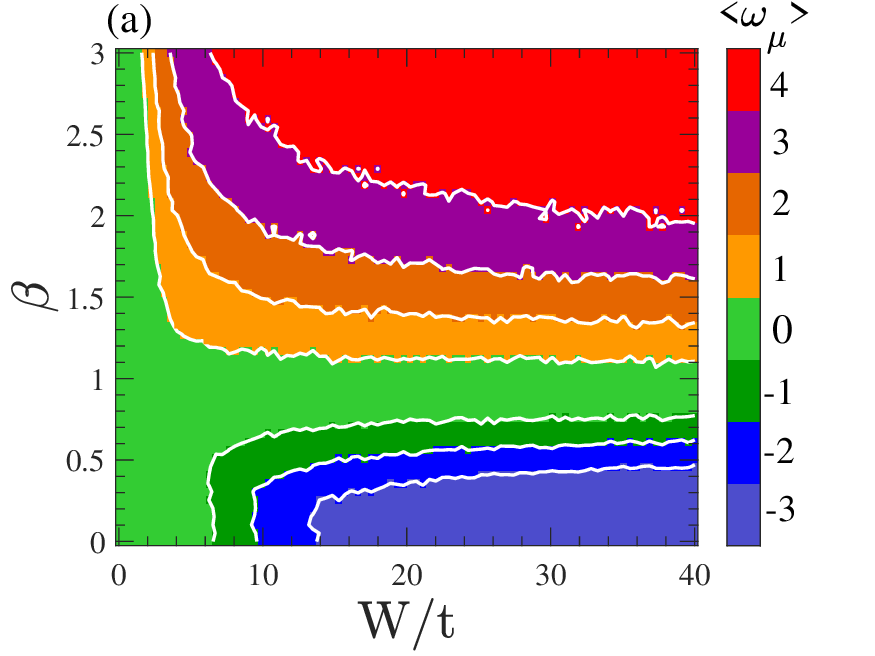}
\includegraphics[width=0.33\columnwidth]{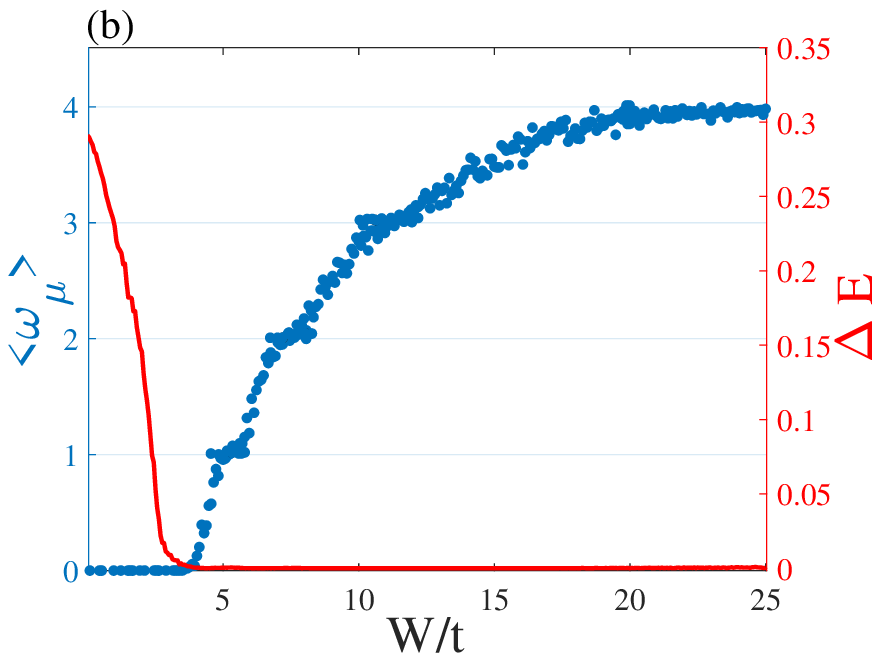}
\includegraphics[width=0.33\columnwidth]{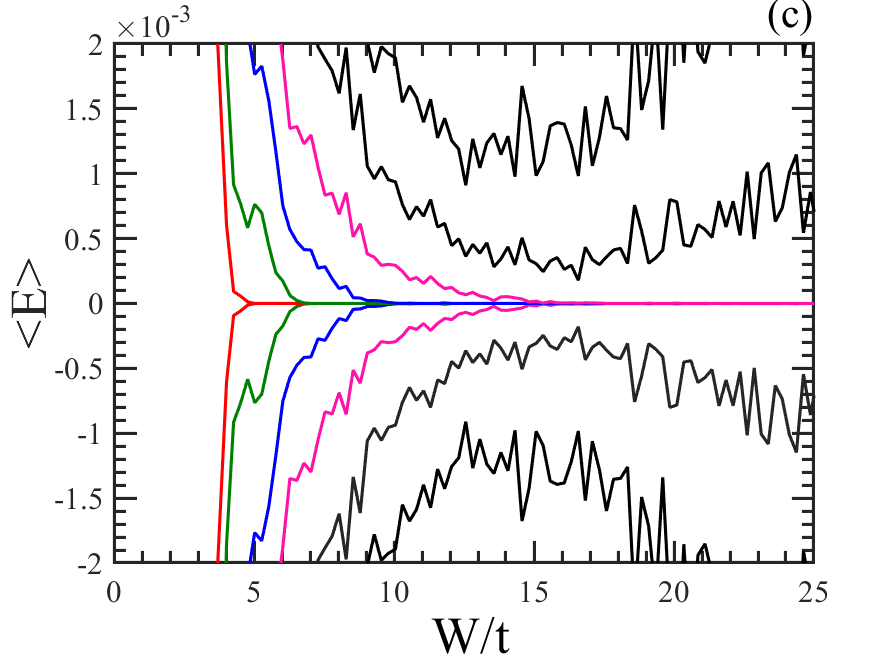}
\end{array}$

\includegraphics[width=0.95\columnwidth]{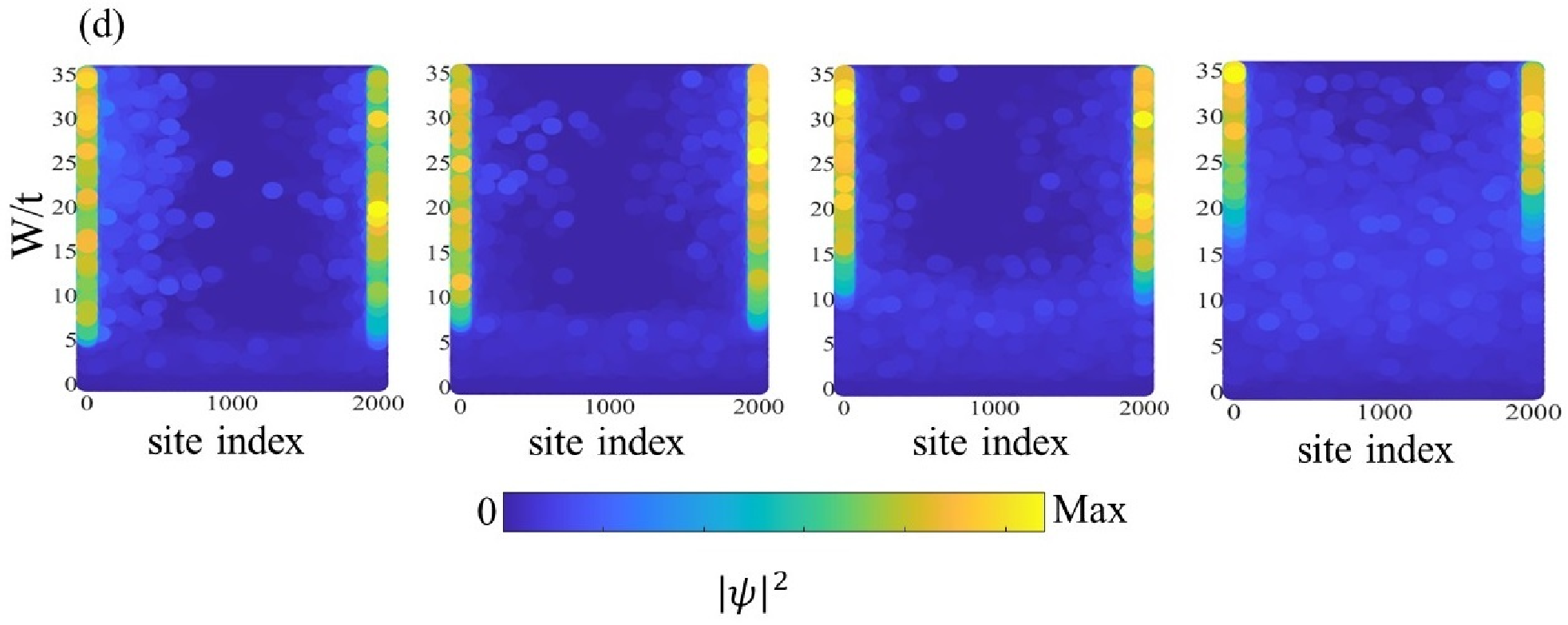} 

\vspace{-2mm} 
\caption{\textbf{Robustness of topological Anderson insulating phases in SWNT.} (a) Topological phase diagram as a function of the disorder amplitude $W$ and the asymmetry parameter $\beta$ (Eq.~\ref{J2J3}) for a $(8, 6)$-SWNT with $L=600$ unit cells. The other parameter are $p = q = -1,$ and $\mu = 0$. (b) Real-space winding numbers $\omega_\mu$ and bulk gap $(\Delta E)$ versus $W$ for the same nanotube with size $L=10^3$ unit cells and $\beta=2.8$. (c) Corresponding disorder-averaged spectra of finite 1D SWNT chain with open boundary conditions as a function of $W$ for the same parameters sets. The different colored lines indicate edge states, while black lines correspond to bulk states. (e) Probability-density maps of the midgap modes $\psi_N$, $\psi_{N-1}$, $\psi_{N-2}$, and $\psi_{N-3}$ as a function of disorder strength $W$. In our numerical calculations, an average over $50$ disorder realizations was performed.}
\label{PhaseDiagDiso}
\end{figure}

\smallskip 
In Fig. \ref{PhaseDiagDiso}(a), we present the disorder-averaged topological phase diagram for the $(8,6)$ nanotube in the $(W, \beta)$ plane. Along the line $\beta \approx 1$, where disorder acts symmetrically on $J_2$ and $J_3$, the winding number remains $\langle\omega_\mu\rangle=0$ for all $W$, and no TAI phase appears. By contrast, for $\beta>1$, the disorder on the $J_2$ hoppings dominates, and we observe a staircase of TAI phases in which $\langle\omega_\mu\rangle$ increases as $W$ increases, eventually approaching the extremal value $+n/d$. Likewise, for $\beta<1$, the disordered $J_3$ dominates and an analogous staircase towards negative winding number $-m/d$ is obtained.\\

\smallskip
We also calculate the averaged bulk energy gap $\Delta E$ via the exact diagonalization of the lattice Hamiltonian in Eq.~\eqref{SWNT-offDiagonal} with periodic boundary conditions, defining $\Delta E=E_{N+1}-E_N$ with $E_i$ being the $i$-th eigenvalue. In Fig. \ref{PhaseDiagDiso}(b), we plot the disorder-averaged winding number $\langle\omega_\mu\rangle$ and the bulk spectral gap as a function of the disorder amplitude $W$. In the clean limit and in the weak disorder regime, the system is in the trivial insulating state with $\langle\omega_\mu\rangle =0$ and a finite bulk gap. Upon increasing $W$, a first topological transition occurs at critical disorders $W_{c} \approx 3.4 t$, where the bulk spectral gap closes and $\langle\omega_\mu\rangle$ jumps to a plateau at $1$, marking the emergence of a TAI phase. This initial transition can be described within a self-consistent Born approximation~\cite{Beenakker}, which captures the disorder-induced renormalization of long-range intercell hopping and the associated closing of the bulk gap. A detailed analysis of the analytical results obtained by this approximation is presented in the Supplementary Note 4. For larger $W$, the bulk spectrum remains gapless while the quantized value of $\langle\omega_\mu\rangle$ evolves stepwise up to $+n/d$. These findings demonstrate that the sequence of TAI transitions are controlled by a mobility gap rather than a simple reopening of the spectral bulk gap, as we substantiate in the following using open-boundary spectra and density of states analysis. Finally, in contrast to previous models \cite{UngappedTAI,ungappedTAI1,Successive,Successive2}, further increasing $W$ does not drive the system back to a trivial Anderson insulator since $\langle\omega_\mu\rangle$ does not vanish and remains at a plateau.

\smallskip 
To elucidate the mechanism behind this phenomenon, we calculate the energy spectrum of an open chain as a function of disorder, as shown in Fig. \ref{PhaseDiagDiso}(c). As described above, for low disorder, the system is trivial without zero-energy states. Beyond the critical value $W_{c}$, the $N$-th and $(N+1)$-th eigenstates become degenerate at $E=0$. By increasing the disorder strength, $|\langle\omega_\mu\rangle|$ pairs of eigenmodes converge towards zero energy in close succession, according to the bulk-edge correspondence. For the highest plateau, we find four pairs of zero- modes (e.g. $\psi_{N-3}, \cdots, \psi_{N+4}$), consistent with $\langle\omega_\mu\rangle =4$. 

\smallskip 
These zero-energy modes reside within a finite mobility subgap $\Delta E_M$, defined as the interval between $E=0$ and the nearest bulk eigenvalues, namely $\Delta E_M= E_{N+5}-E_{N-4}$. Figure \ref{PhaseDiagDiso}(c) shows that the mobility subgap does not vanish; instead, it grows as disorder increases. Therefore, the degenerate zero-energy states stay well-separated from the bulk states and remain highly robust, without splitting even at strong disorder. Consequently, the system will remain in a TAI phase throughout the plateau. In Supplementary Note 5, we further characterize this mobility gap using the ratio of the geometric mean DOS $\rho_{\text{typ}} $ to the arithmetic mean DOS $\rho_{\text{ave}}$ at $E=0$~\cite{arithmeticDOS1,arithmeticDOS2}. Moreover, we analyze the disorder-averaged spatial profiles of the midgap states, denoted by the colored lines in Fig.~\ref{PhaseDiagDiso}(c), as a function of disorder amplitude $W$. Figure \ref{PhaseDiagDiso}(d) shows that for weak disorder, these eigenstates are quite extended in the bulk and become highly localized at the two ends of the chain successively as the disorder increases.

\smallskip
For comparison, it is worth noting that we have also studied the case where disorder acts on the intracell hopping $J_1$ rather than on the long-range bonds. In this case, the system exhibits the conventional 1D TAI behavior, with a single transition to a phase with $\langle\omega_\mu\rangle =1$ followed by a trivial Anderson insulator at strong disorder (see Supplementary Note 6). This result highlights that the robust staircase reported here relies crucially on disorder in the long-range couplings. 
\subsection{Localization properties}
To further confirm the staircase phase transitions, we compute the localization length $\Lambda$ of the midgap modes using the transfer matrix method \cite{LEdefinition1,LEdefinition2} (see Methods section for details). Generally, a divergence in $\Lambda$ characterizes the delocalized states that appear at transition points between different phases in 1D chiral chains \cite{Prodan,LLtransition}. We choose a long system length $N$ to guarantee the numerical convergence.

\begin{figure}[hpbt] 
$\begin{array} {cc}
\includegraphics[width=0.4\columnwidth]{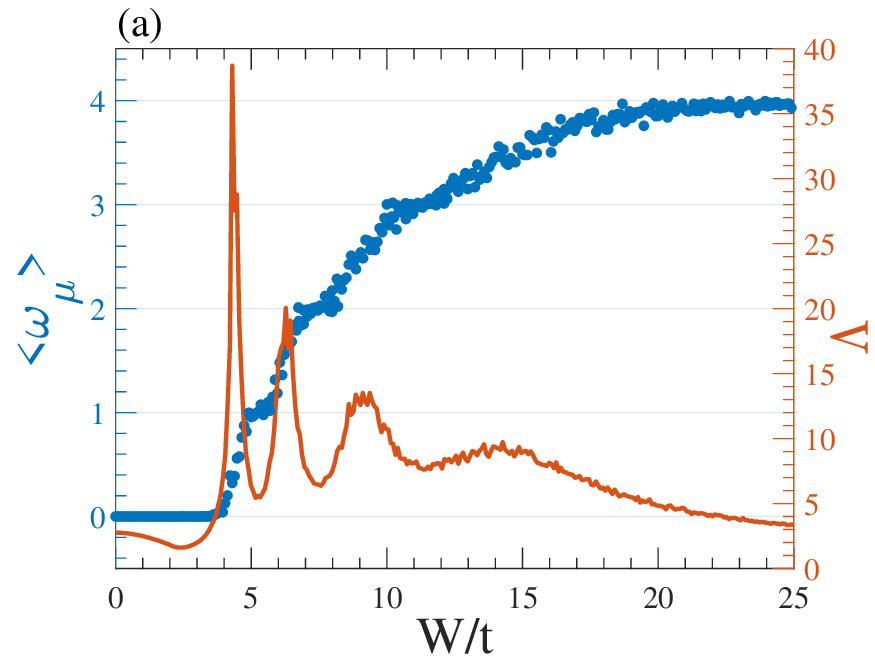}
\includegraphics[width=0.4\columnwidth]{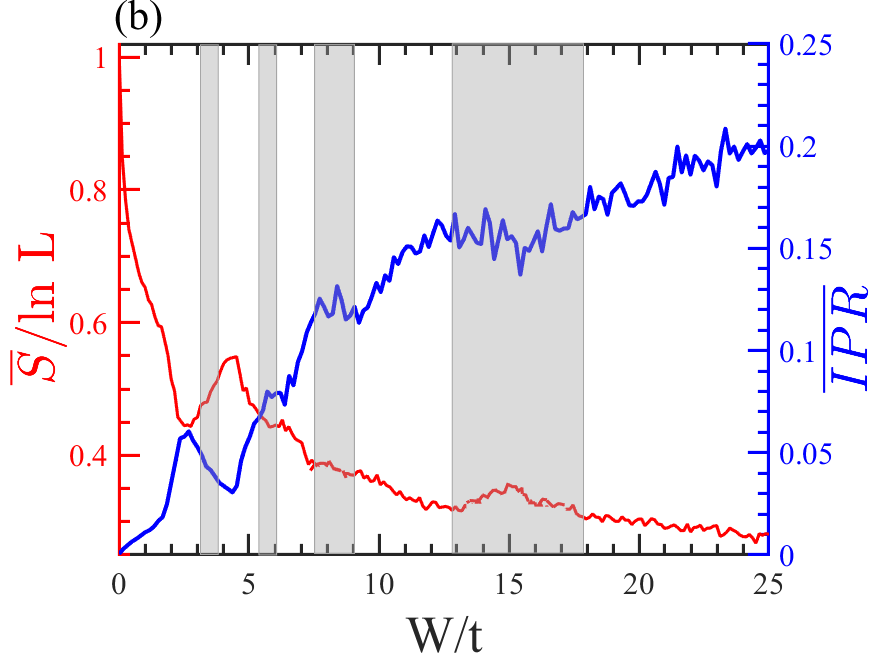}
\end{array}$
\vspace{-2mm} 
\caption{\textbf{Localization properties of midgap eigenstates.} (a) Average winding number and localization length at $E=0$ as a function of disorder strength $W$ for a $(8,6)$-SWNT with $p=q=-1$, $\mu=0$. (b) Average $\overline{\text{IPR}}$ and mean Shannon entropy $\overline{\text{S}}/\ln{L}$ of the midgap states as a function of $W$. Other parameters are the same as those in Fig. \ref{PhaseDiagDiso}.}
\label{AverageIPR}
\end{figure}

\smallskip 
The results are shown in Fig. \ref{AverageIPR}(a), where we calculate the localization length at the Fermi level $(E=0)$ for the $(8,6)$ SWNTs under open boundary conditions. It is clear that the Anderson localization-delocalization transitions perfectly fit the topological phase boundaries. In the trivial gapped phases, disorder is found to localize all the evanescent wavefunctions present at $E=0$. However, at a topological phase transition, $\langle\omega_\mu\rangle$ loses quantization and $\Lambda$ shows a peak, as expected. Inside the TAI regions, $\Lambda$ decreases with an increase of $W$, indicating that the states around zero energy remain localized. As the transitions move to larger disorder, the $\Lambda$ peaks become smaller and broader, reflecting the suppression of delocalization and the tendency of all states to localize. After the system reaches the TAI with higher winding number $\langle\omega_\mu\rangle = +n/d$, no peak in $\Lambda$ is observed. This feature is corroborated by the behavior of the Lyapunov exponent $\gamma(W)$~\cite{LEdefinition1,LEdefinition2}, which can be derived analytically in the strong disorder limit for $\beta \gg 1$ as (see Supplementary Note 7 for details)
\begin{equation}
    \gamma(W)= \big\langle \ln | J_2| \big\rangle - \ln |t|\,.
\end{equation}
For $W \gg t$, $\gamma(W)$ is always positive, which expresses the robustness of this phase against strong disorder .

\smallskip 
To characterize the localized or extended nature of the mid-gap states, we calculate the average inverse participation ratio (IPR) of the low-energy modes as \cite{IPRdefinition}
\begin{equation}
\overline{\text{IPR}}=\frac{1}{N_E} \sum_{n=1}^{N_E} \text{IPR}_n, \quad \text{with} \quad \text{IPR}_n= \sum_{r,\alpha} |\psi_\alpha^n (\mathbf{r})|^4.
\end{equation}
Here $\psi_\alpha^n (\mathbf{r})$ is the normalized wavefunction of the $n$-th eigenstate at site $\mathbf{r}$ and sublattice $\alpha$, and $N_E$ is the number of the eigenstates closer to $E=0$. In the thermodynamic limit, $L\rightarrow \infty$, the system exhibits extended phases when $\overline{\text{IPR}}$ approaches to zero as $\propto L^{-1}$, whereas for localized states, it tends to a constant.
\smallskip 
In addition, we computed the mean Shannon (information) entropy, defined as \cite{Shanon1,Shanon2}
\begin{equation}
\overline{\text{S}}=\frac{1}{N_E} \sum_{n=1}^{N_E} \text{S}_n, \quad \quad \text{S}_n= -\sum_{j} |\psi_j^n|^2 \ln{\big(|\psi_j^n|^2\big)}.
\end{equation}
This quantity is also an important parameter to clarify the localization properties of the eigenstates. We expect the $\overline{\text{S}}$ of the strongly localized states to vanish while $\overline{\text{S}}=\ln{L}$ for ergodically extended states. The results of the $\overline{\text{IPR}}$ and the renormalized $\overline{\text{S}}/\ln{L}$ are plotted in Fig. \ref{AverageIPR}(b), where the mean values are taken over the first eight eigenstates around $E=0$. Initially, for $W=0$, the extended nature of the midgap modes is clearly shown with $\overline{\text{IPR}}=0$ and $\overline{\text{S}}/\ln{L} \sim \mathcal{O}(1)$. Upon increasing the disorder strength, the $\overline{\text{IPR}}$ becomes finite and the Shannon entropy decreases in the TAI phases, which means that the states around $E=0$ are localized at the chain ends. The sharp dips in the $\overline{\text{IPR}}$ curve and the anomalous peaks in the $\overline{\text{S}}$ curve provide convincing evidence for the series of localization-delocalization transitions. Importantly, unlike the monotonic increase behavior of a trivial Anderson insulator, $\overline{\text{IPR}}$ remains constant in the strong disorder regime. 
\subsection{Experimental validation}
\begin{figure}[hpbt] 
    \centering
    \includegraphics[width=0.9\linewidth]{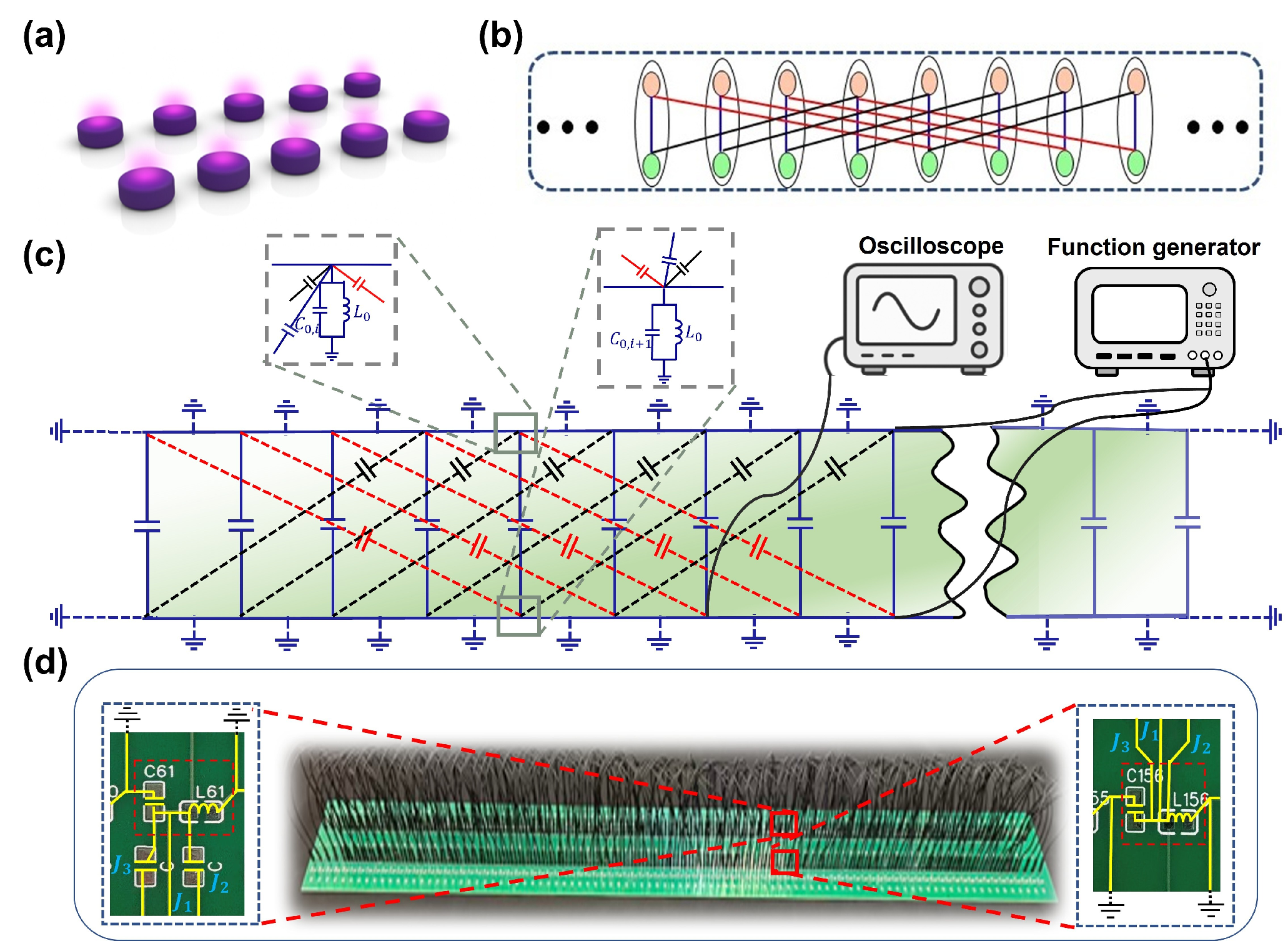}
    \vspace{-2mm} 
    \caption{\textbf{Experimental topolectrical circuit realization of a robust TAI cascade in the reduced 1D SWNT model.} (a) Artistic view of the two coupled 1D chains of resonant nodes representing the A and B sublattices of the effective 1D SWNT model. (b) Reduced 1D model of the $(8,6)$-SWNT. Red lines denote the forward long-range coupling $J_2$, black lines the backward long-range coupling $J_3$, and blue lines the direct intracell coupling $J_1$ between two sites within the same unit cell. (c) Top view of the equivalent 1D topolectrical circuit, together with enlarged views highlighting the resonator sites and grounding elements. A function generator provides the excitation signal at selected nodes, and a digital oscilloscope is used to detect the resulting voltage response, as schematically indicated on the right side. (d) Photograph of the experimental setup with a size of 2 × 95 nodes. The zoomed-in views show the detailed structure of two unit cells.
}
    \label{Setup}
\end{figure}

In light of recent experimental progress in engineering a variety of quantum phases with circuit-based platforms \cite{Circuit,Circuit1,Circuit2,Circuit3,Circuit4,Circuit5,Circuit6,Circuit7}, we proceed to design 1D topolectrical circuits with disordered long-range amplitudes that can host the topological states described above. Figure \ref{Setup}(a) gives an artistic view of two coupled 1D chains of resonant nodes that represent the two electronic sublattices of the effective 1D SWNT model. The intersite circuit couplings realize the tight-binding hopping of the theoretical Hamiltonian. The reduced 1D model of the $(8,6)$-SWNT is shown schematically in Fig. \ref{Setup}(b). The uniform intracell coupling $J_1=t$ between two sites in the same unit cell is indicated by a blue line, while the forward and backward long-range coupling $J_2$ and $J_3$ between distant unit cells are indicated by red and black lines, respectively.

\smallskip 

\smallskip
Figure \ref{Setup}(c) shows the equivalent topolectrical circuit that maps the tight-binding Hamiltonian of Eq. \eqref{SWNT-offDiagonal} onto an admittance network governed by Kirchhoff's law. Each lattice site is represented by a resonant node consisting of a parallel $L_0-C_0$ tank, and the couplings between sites are realized by capacitors of different values that emulate the various hopping amplitudes. Disorder is programmed only in the long-range intercell couplings. In practice, we randomize the coupling capacitors associated with $J_2$ and $J_3$ around their nominal values according to a prescribed distribution. Thus, disordered hoppings are implemented by capacitors with different capacitance values that emulate different hopping amplitudes between distant sites. For each disorder strength $W$, we implement and characterize multiple statistically independent disorder realizations.

The grounding capacitor $C_0$ at each node is a local tuning parameter that compensates the disordered coupling capacitances, ensuring that the total capacitance per node, and hence the resonance frequency, is uniform throughout the circuit. Only inter-sublattice links between the A and B nodes are implemented, while intra-sublattice links are omitted, preserving the chiral symmetry of the theoretical model. To ensure a uniform on-site term, we also compensate the missing branch admittance at the boundary nodes by suitable grounding elements so that all nodes share the same resonance angular frequency $\nu_0=1/\sqrt{L_0C_{\text{tot}}}$, where $C_{\text{tot}}$ denotes the common total capacitance per node. As a consequence, the circuit Laplacian $J(\nu)$ faithfully reproduces the off-diagonal tight-binding model, with the zero-energy point mapped to the resonance frequency $\nu_0$. 

\smallskip
In Fig. \ref{Setup}(d), a photograph of the fabricated circuit is shown with $L=95$ unit cells ($190$ nodes), which is large enough to ensure convergence of our observables with respect to finite-size effects. The insets provide enlarged views of two neighboring unit cells, showing the detailed configuration of the capacitors and inductors. A detailed description of the sample fabrication and experimental setup is provided in the Methods section.

\smallskip 
In the linear AC regime, voltage measurements provide a direct probe of the spectrum of the circuit Laplacian. At a fixed angular frequency $\nu$, Kirchoff's law states that the measured voltage profile is expressed as \cite{Voltagedefinition}
\begin{equation}
    V(\nu)=\sum_j \frac{\braket{\psi_j|I(\nu)}}{\epsilon_j(\nu)} \ket{\psi_j},
\end{equation}
where $\epsilon_j(\nu)$ and $\ket{\psi_j}$ are the eigenvalues and eigenmodes of $J(\nu)$, and $I(\nu)$ is the vector of injected currents. This expression shows that an edge mode whose Laplacian eigenvalue is nearly zero at a midgap frequency $\nu^*$ dominates the measured voltage profile.

\smallskip 
\begin{figure}[hpbt] 
    \centering
    \includegraphics[width=0.9\linewidth]{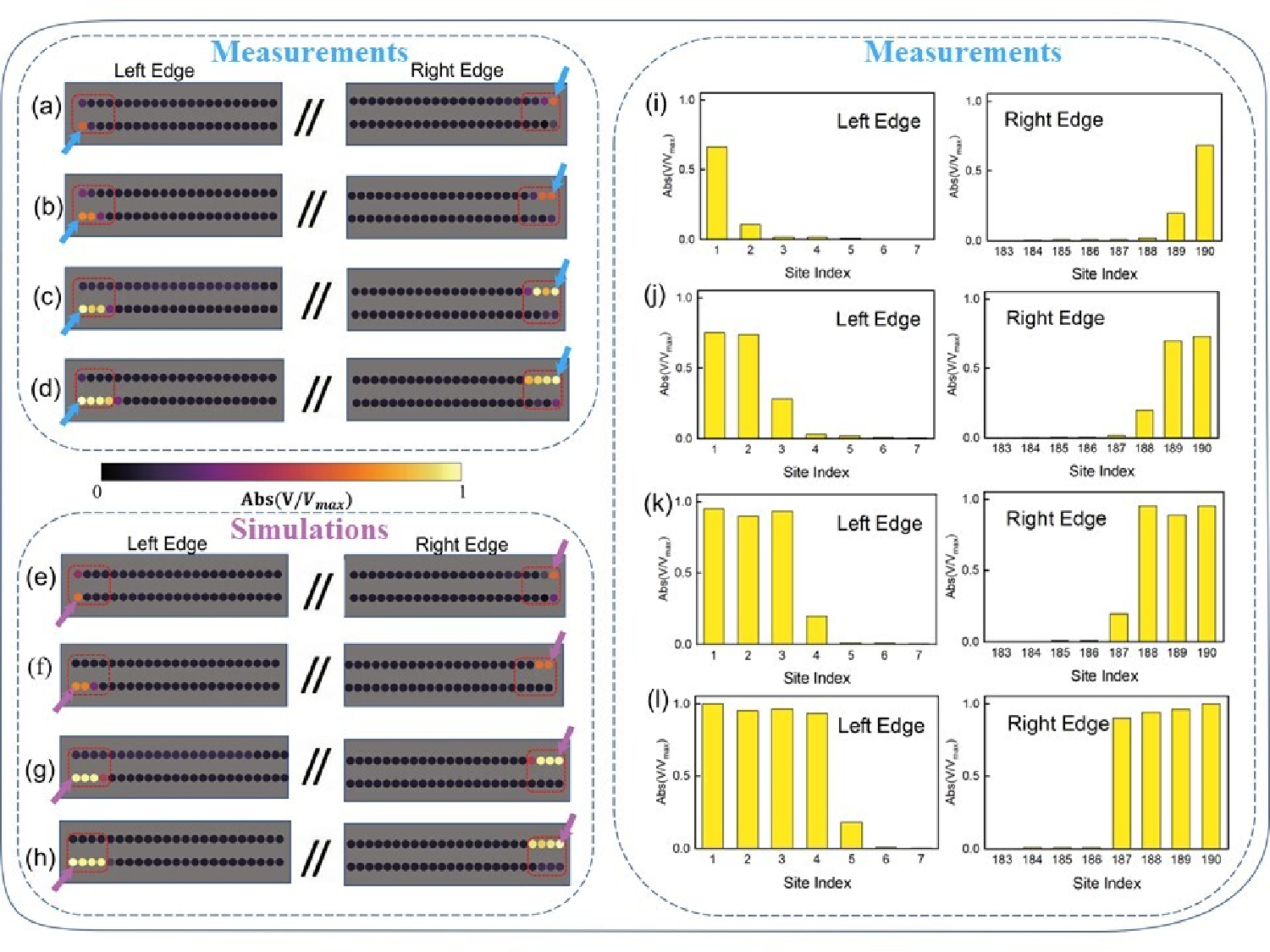}
    \caption{\textbf{Observation of successive emergence of disorder-induced edge modes in the topolectrical SWNT analogue.} (a-d) Measured normalized voltage distributions at the edge-state resonance frequency for disorder strengths $W= 7.8t,\ 10.9t,\ 14.6t,$ and $22t$, respectively. (e-h) Corresponding LTspice-simulated voltage profiles for the same disorder strengths and drive conditions. The dashed red box highlights the edge area selected for display. The blue and purple arrows mark the excitation regions of the signal excitation. (i-l) Histogram representation of the experimental edge-state voltage profiles. The noramlized amplitude of $|V_j/V_{\text{max}}|$ is shown as a bar plot versus site index for disorder amplitudes corresponding to (a-d) cases.
}
    \label{Measurement}
\end{figure}

\smallskip
To probe the edge states experimentally, we excite a single boundary node with a sinusoidal voltage and measure the spatial voltage distribution at the resonance frequency $f_0=\nu_0/2\pi$, as shown in Figs. \ref{Measurement}(a)
and \ref{Measurement}(d). The normalized amplitude profile exhibits large values at the two ends of the chain and decays rapidly into the bulk, directly revealing an edge-localized mode. As the disorder strength $W$ is increased, we observe the predicted disorder-driven staircase of topological transitions where the number of edge modes grows in discrete steps. At weak disorder, a single pair of edge peaks indicates a topological phase with winding number $\langle\omega_\mu\rangle =1$. By increasing $W$ again, additional pairs of edge peaks emerge further, reaching four protected edge resonances (Fig~\ref{Measurement}(d)). Figures \ref{Measurement}(e-h) show the corresponding AC simulations obtained from LTspice software using the same parameters. The simulated node-voltage profiles closely match the measured ones, confirming that the topolectrical circuit realizes the reduced SWNT and that the observed responses correspond to disorder-induced midgap edge states.

\smallskip
To check the number of edge states emerging at the ends of the chain,  we plot, in Figs.~\ref{Measurement}(i-l), the measured voltage values $|V_j/V_{\text{max}}|$ as bar plots along the chain at the edge-state resonance frequency $f_0$ and for different disorder amplitudes corresponding to those used in Figs.~\ref{Measurement}(a-d). At weak disorder (Fig.~\ref{Measurement}(i)), a single dominant peak is localized near each chain boundary. By increasing disorder, 
 well-separated peaks appear whose voltages cannot be fitted by a decaying exponential (Fig.~\ref{Measurement}(j-l)), which indicates an increasing number of distinct topological edge states, rather than a single edge mode progressively extending into the bulk.
\section{Conclusion}
We have proposed a model for step-like topological Anderson phase transitions in a reduced SWNT with chiral-symmetry-preserving disorder. We have implemented the model in an electric circuit where node-voltage measurements and LTspice simulations clearly show the emergence of edge states with a stepped increasing number. Unlike the conventional picture of TAI, where the system becomes a trivial Anderson insulator at sufficiently large disorder, our system remains in a topological phase. By introducing disorder only on the long-range intercell hoppings, we have shown that the mobility subgap around $E = 0$ does not collapse and that the edge modes persist up a large disorder amplitude. We have also corroborated this finding by examining the localization length, participation ratios, and Shannon entropy of the midgap eigenstates. Our work extends the TAI paradigm and paves the way to designing platforms with disorder-tuned topology. 
 \section{Methods}
 \subsection{Transfer matrix method for model with long-range hopping}
Recently, the TMM was extended to calculate the localization length of a disordered 
tight-binding Hamiltonian with arbitrary long-range hopping terms \cite{GeneralizedTMM1,GeneralizedTMM2}. For our model, each unit cell $l$ has forward off-diagonal hopping terms to cell $l+n/d$ and backward ones to cell $l-m/d$. To prevent hopping from extending beyond the nearest neighbor's supercell, we redefine the original unit cells into supercells of length 
\begin{equation}
    \delta_{max}=\frac{n+m}{d}+1.
\end{equation}
Letting $\Psi_i=(u_{A,l}, u_{B,l}, \cdots, u_{A,l+\delta_{max}-1}, u_{B,l+\delta_{max}-1})^T$ be the amplitude of the wavefunction at supercell $i$, we can write the Schr\"{o}dinger equation for a given energy $E$ as
\begin{equation}
    E \Psi_i= J_{i,i+1} \Psi_{i+1}+M_i \Psi_i+ J_{i,i-1} \Psi_{i-1},
\end{equation}
where $M_i$ is a $(2\delta_{max} \times 2\delta_{max}) $ matrix corresponding to intracell hopping in the supercell $i$, and $J_{i, i\pm 1}$ is the coupling matrix containing all long-range hoppings between the $i$th supercell and the $(i\pm 1)$th supercell.

\smallskip 
At zero energy, the recursion relation yields  the nearest-neighbor transfer matrix 
\begin{equation}
    \begin{pmatrix} \Psi_{i+1} \\ \Psi_i \end{pmatrix} = T_i \begin{pmatrix} \Psi_i \\ \Psi_{i-1} \end{pmatrix}, \quad T_i = \begin{pmatrix} - J_{i+1}^{-1} M_i & - J_{i+1}^{-1} J_{i-1} \\ 1_{(2\delta_{max}\times2\delta_{max})} & 0_{(2\delta_{max}\times2\delta_{max})} \end{pmatrix}.
\end{equation}
The consecutive product of transfer matrices, $\tau_N=\prod_{i=0}^N T_i$, provides information on the evolution of the zero-energy modes. The limit $lim_{N \rightarrow \infty} (\tau_N^{\dagger}\tau_N)^{(1/2N)}$ has positive eigenvalues of the form $\exp{(\pm \gamma_i)}$, where $\gamma_i$ are so-called Lyapunov exponents \cite{LEdefinition1,LEdefinition2}. The localization length is nothing but the inverse of the smallest $\gamma_i$ 
\begin{equation}
    \Lambda=\frac{1}{\gamma_{\text{min}}}\,.
\end{equation}
\subsection{Sample fabrications and circuit measurements}
The experimental circuit is constructed on a printed circuit board with a thickness of $1.6$ mm. The electronic components include surface-mount capacitors and inductors. Each inductor has an inductance of {$L_{0}= 2.2\,\text{uH}$}. The total capacitance at each node is set to $C_{\text{tot}}= 13 \text{nF}$, ensuring identical on-site terms. Therefore, the resonance frequency $f_0$ of each node is given by $f_0=1/2\pi \sqrt{L_{0}C_{\text{tot}}}=941.58\,\text{kHz}$. To reduce the disorder caused by the tolerances of capacitors and inductors, we manually selected the capacitors and inductor values by using an impedance analyzer (Keysight E4990A) and controlled their tolerances within 2\%. When designing the circuit, it is crucial to maintain an appropriate spacing between components to avoid short circuits. Additionally, the power supply traces should be as wide as possible to prevent excessive current from causing overheating or damage. In our experimental setup, jumper wires are used to connect certain circuit components, thereby introducing disorder into the long-range intercell hopping, as depicted in Fig. \ref{Setup}.

\smallskip 
For the measurements, a function generator (KEYSIGHT 33600A) provides small-signal AC excitation at selected edge nodes with a frequency of $941.58\,\text{kHz}$, and the oscilloscope (SIGLENT SDS1202X-E) is used to capture the corresponding voltage response. Initially, the excitation signal is verified using an oscilloscope. Once confirmed to be accurate, the response voltage signals at the required nodes are measured and recorded sequentially. The voltage distribution at the corresponding states is then characterized by the node response voltage distribution at the corresponding eigenfrequency. The function generator used in the experiment has an internal resistance of $50 \Omega$. As shown in Fig. \ref{Measurement}, the blue and purple arrows indicate the excitation regions of the signal. We measured the voltage amplitude $|V_j( \omega_0)|$ at accessible nodes. The spatial voltage distributions and normalized amplitude maps shown in the main text are constructed from these measured node-voltage amplitudes.  
\section{Data availability}
The numerical data for generating the figures are available from the authors upon reasonable
request.
\section{Code availability}
The numerical codes for generating the results are also available from the authors upon request.
\section*{Acknowledgements}
We are grateful to Andrej Mesaros, Jean-No\"el Fuchs and Fr\'ed\'eric Pi\'echon for fruitful discussions and a critical reading of the manuscript. We also thank Udo Schwingenschl\"ogl for interesting discussions.
M.M., H.S. acknowledge support by Tamkeen under the NYU Abu Dhabi Research Institute grant CG008. M.M., H.S. acknowledge further support from the NYU IT High Performance Computing services. S.Y., R.M., C.H., S.Y. acknowledge support by the National Key Research Program of China (Grant No. 2021YFA1400602), and by the National Natural Science Foundation of China (Grant No. 12474316).  S.H acknowledges CQTS at NYU Abu Dhabi for supporting her visit during which this project was initiated.
\section*{Competing Interests}
The authors declare no competing interests.
\section*{Author Contributions statement} 
M.M and S.Y. contributed equally to this work. M.M. performed the analytical and the numerical calculations. S.Y., R.M. and Y.S. realized the experimental set-up and performed the measurements. S.H. and M.M. conceived the idea and discussed the different steps. S.H., Y.S. and H.S. guided the research. All the authors contributed to the discussions of the results and the preparation of the manuscript.
\section*{Supplementary Information for: Realization of staircase topological Anderson phase transitions}

\begin{itemize}
\item \hyperref[Note1]{Supplementary Note 1: Clean effective Hamiltonian}\\
\item\hyperref[Note2]{Supplementary Note 2: Semiconducting SWNTs with $n = m + 1$}\\
\item\hyperref[Note3]{Supplementary Note 3: Finite-size scaling}\\
\item\hyperref[Note4]{Supplementary Note 4: Self-consistent Born approximation}\\
\item\hyperref[Note5]{Supplementary Note 5: Arithmetic and geometric mean DOS}\\
\item\hyperref[Note6] {Supplementary Note 6: Intracell-disorder case}\\
\item\hyperref[Note7]{Supplementary Note 7: Persistence of higher winding number topological Anderson phases}\\

\end{itemize}

\vspace*{\fill}
\newpage
\subsection{Supplementary Note 1: Clean effective Hamiltonian}\label{Note1}
\begin{figure}[hpbt] 
\includegraphics[width=0.5\columnwidth]{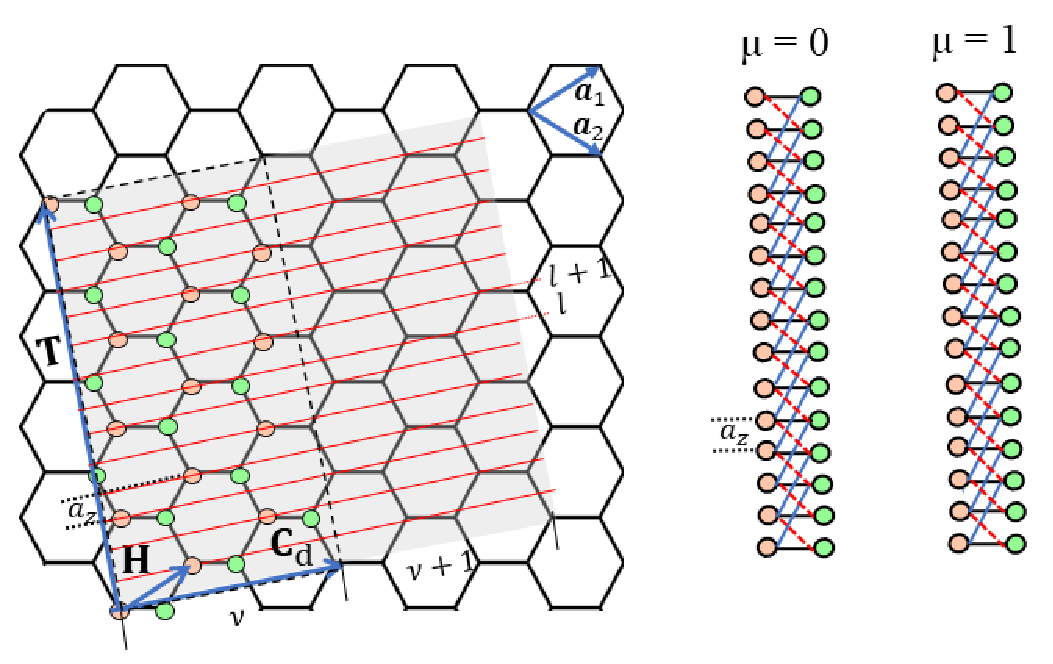}
\vspace{-3mm} 
    \caption{(a) Unrolled two-dimensional hexagonal lattice of the $(4,2)$ SWNT. Wrapping the chiral vector $\mathbf{C}= 4 \mathbf{a_1} + 2 \mathbf{a_2}$ onto itself results in the nanotube with axis along a translation vector $\mathbf{T}$. The elementary vector $\mathbf{C}_d\equiv\mathbf{C}/d$ corresponds to a rotation of $\frac{2\pi}{d}$ around the tubule axis. The dashed rectangle delimits the unit cell used to define the reduced 1D 
    model, while the grey one between $\mathbf{C}$ and $\mathbf{T}$ is the larger translational unit cell. In this case $(p,q)=(1,0)$ and $d=gcd(m,n)=2$. (b) Schematic illustration of the corresponding two 1D chains. The black solid lines denote the intracell hopping $J_1$, while the blue lines and red dashed lines indicate the $\mu$-dependent long-range hopping parameters $J_{2,\mu}$ and $J_{3,\mu}$, respectively.}
    \label{Shematic}
\end{figure}

To get insights on the topological properties of $(n,m)$-SWNTs, a helical-angular description based on their translational and rotational symmetries is often used \cite{topologyCNT2,topologyCNT3}. In this framework, a 1D reduced model is obtained by introducing a helical vector $\mathbf{H}= p \mathbf{a_1} + q \mathbf{a_2}$ such that the planar graphene sheet can be mapped into 
$d={\rm gcd}(n,m)$ spiral chains parallel to $\mathbf{H}$. The lattice sites can then be labeled as 
\begin{equation}
     \mathbf{r}_{\nu,l}= \nu \frac{\mathbf{C}}{d} + l \mathbf{H}, \quad \nu = 0,1,\dots,d-1,\ l \in \mathbb{Z}\,,
\end{equation}
where $\nu$ labels the helical strand and $l$ indicates the position along $\mathbf{H}$ in units of $a_z$ where $a_z = \sqrt{3}d a^2/2 \pi d_t$ and $d_t$ is the nanotube diameter $d_t=|\mathbf{C}|/\pi$. The vector $\mathbf{C}/d$ corresponds to a rotation by $2\pi/d$ around the tube axis, so that the angular momentum $\mu \in \left\{0, ..., d-1 \right\}$ associated with translations by $\mathbf{C}/d$ is a good quantum number. For each fixed $\mu$, the tight-binding Hamiltonian reduces to an effective 1D system with $N$ unit cells, each containing two sites. The vectors $\mathbf{C}/d$ and $\mathbf{H}$ thus form the elementary cell in the helical-angular construction. A schematic example for a $(4,2)$-SWNT is shown in Fig. \ref{Shematic}(a). The corresponding two 1D chains (for $\mu=0,\ 1$ in this example) are presented in Fig. \ref{Shematic}(b), with $\mu$-dependent long-range hoppings represented by colored lines. 

In the clean limit, $W=0$, we can decompose the Bloch Hamiltonian as 
\begin{equation} \label{cleanH}
    H_{0,\mu}(k)= \mathbf{d}_\mu(k) \mathbf{\sigma}\,,
\end{equation}
where 
\begin{align} \label{componentsd}
    d_{x,\mu} &= t\bigg(1 + J_{2,\mu} \cos\Big(\frac{kn}{d}\Big) 
    + J_{3,\mu} \cos\Big(\frac{km}{d}\Big)\!\bigg),
    \\
    d_{y,\mu} &= t\bigg(J_{2,\mu} \sin\Big(\frac{kn}{d}\Big)
    - J_{3,\mu} \sin\Big(\frac{km}{d}\Big)\!\bigg),
\end{align}
$\sigma_\alpha (\alpha =x,y)$ are the Pauli matrices describing the two sublattices, and $k= \mathbf{k} \cdot \mathbf{H}$ is the Bloch momentum along $\mathbf{H}$. The values $(n,m,p,q, \mu)$ are chosen such that the phase factors 
\begin{equation}
    J_{2,\mu}=e^{\frac{-i 2\pi \mu p}{d}} \quad \text{and} \quad J_{3,\mu}=e^{\frac{i2\pi \mu q}{d}}
\end{equation} are real numbers. The corresponding dispersion relation is $ E_{0,\mu}(k)= \pm t\sqrt{d_{x,\mu}^2+d_{y,\mu}^2}$. In the semiconducting case, $\text{mod}(2n+m, 3) \neq 0$, the conduction and valence bands are separated by a gap, and therefore the chain is insulating at half-filling.

\smallskip 
The reduced 1D-SWNT Hamiltonian given by Eq. \eqref{cleanH} is time-reversal symmetric with $T^2=+1$ because the hopping parameters are real. Chiral and particle-hole symmetries are also preserved since $\sigma_z H_{0,\mu}(k) \sigma_z = - H_{0,\mu}(k)$ and $\sigma_z H_{0,\mu}(k)^* \sigma_z = - H_{0,\mu}(-k)$. As a consequence, for each $\mu$, the model belongs to the BDI class of symmetry. Its non-trivial topology is characterized by an integer winding number $\omega_\mu$, which is determined by the chiral indices $(n,m)$ and the angular momentum $\mu$, as summarized in Table \ref{tab-omega_mu} \cite{topologyCNT3}.
\begin{table}[h]
    \centering
    \renewcommand{\arraystretch}{1.5}
    \begin{tabular}{ccc}
        \hline\hline
        \textbf{Type} & \(\omega_\mu\) & \(\mu\) Range \\
        \hline 
        \multirow{2}{*}{\(\text{mod} \left(\frac{2n+m}{d}, 3\right) = 1\)} 
        & \( \frac{(n-m)/d - 2}{3} \) & \( \frac{d}{3} \leq \mu \leq \frac{2d}{3} \) \\
        & \( \frac{(n-m)/d + 1}{3} \) & \( 0 \leq \mu < \frac{d}{3} \text{ or } \frac{2d}{3} < \mu < d \) \\
        \hline
        \multirow{2}{*}{\(\text{mod} \left(\frac{2n+m}{d}, 3\right) = 2\) }
        & \( \frac{(n-m)/d + 2}{3} \) & \( \frac{d}{3} < \mu < \frac{2d}{3} \) \\
        & \( \frac{(n-m)/d - 1}{3} \) & \( 0 \leq \mu \leq \frac{d}{3} \text{ or } \frac{2d}{3} \leq \mu < d \) \\
        \hline \hline
    \end{tabular}
    \caption{Summary of the winding-number values \(\omega_\mu\) for semiconducting SWNTs, where \(\mu\) denotes the angular momentum index.}
    \label{tab-omega_mu}
\end{table}
\\
Most semiconducting SWNTs have been shown to behave as topological insulators with a non-zero winding number, except in the case $n=m+1$ \cite{topologyCNT2,topologyCNT3}.

\begin{figure}[hpbt] 
$\begin{array} {cc}
\includegraphics[width=0.45\columnwidth]{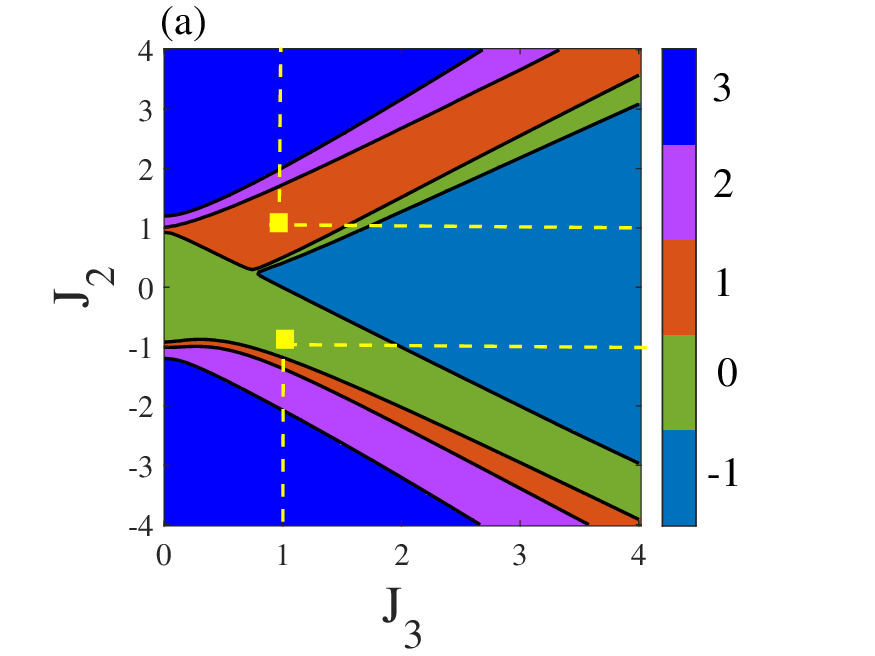}
\includegraphics[width=0.45\columnwidth]{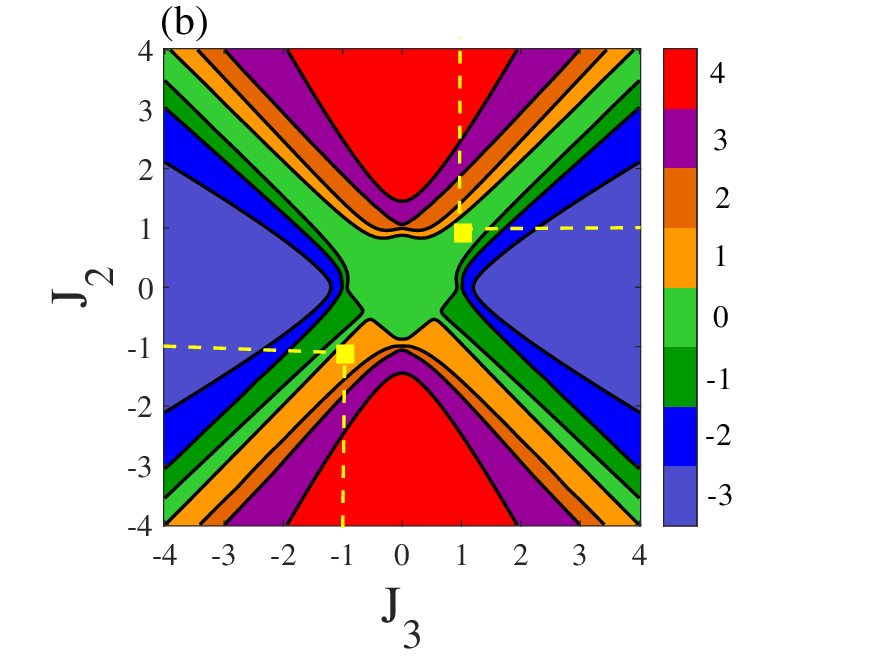}
\end{array}$
\vspace{-3mm} 
    \caption{Map of the winding number $\omega_\mu$ as a function of the intercell hoppings $(J_2, J_3)$ for (a) a type-1 $(12,4)$ semiconducting SWNT and (b) a type-2 $(8,6)$ semiconducting SWNT. The yellow dots mark the $(J_2, J_3)$ values of the clean nanotube for the corresponding angular-momentum $\mu$. The dashed line indicates a representative path along which $|J_2|$ or $|J_3|$ is increased. Along this path, the winding number $\omega_\mu$ undergoes successive topological transitions with stepwise changes, approaching the extremal values $\omega_\mu= +n/d$ and $\omega_\mu= -m/d$ (with $d=\mathrm{gcd}(n,m)$)}.
    \label{Mapwinidng}
\end{figure}

\smallskip 
Figures \ref{Mapwinidng}(a) and \ref{Mapwinidng}(b) represent the winding number as a function of the generalized hopping amplitudes $(J_2, J_3)$ for two representative semiconducting nanotubes. For a type-1 (12,4) SWNT, the angular momenta that produce purely real intercell hoppings are $\mu=0$ and $\mu=2$, which correspond to $(J_2, J_3) = (1,1)$ and $(J_2, J_3) = (-1,1)$, respectively. A type-2 (8,6)-SWNT has only two possible angular momenta, $\mu=0$ and $\mu=1$, which map to intercell hoppings $(J_2, J_3) = (1,1)$ and $(-1,-1)$ respectively. These points lie within regions of $\omega_0=0$ and $\omega_1=1$, as shown by the yellow dots in Fig. \ref{Mapwinidng}(b). As the magnitude of either $J_2$ or $J_3$ increases along the dashed lines, both systems undergo successive topological phase transitions, in which the winding number changes in quantized steps.

\smallskip 
Since we are interested in disorder-induced topology, in the main text we focus on those helical chains whose clean limit is trivial, namely the chain with angular-momentum $\mu=2$ for the $(12,4)$ nanotube and $\mu=0$ for the $(8,6)$ nanotube.

\smallskip 
The extremal values of winding numbers emerge when one of the long-range couplings dominates in the parameter space. When $|J_2| \gg 1+ |J_3|$, the Bloch vector $\mathbf{d}_\mu(k)$ (see Eqs. \eqref{cleanH} and \eqref{componentsd}) can be approximated by $tJ_{2,\mu} e^{i \frac{kn}{d}}$. This simplified form represents a counterclockwise rotation in the complex plane with a phase angle $\theta=\frac{kn}{d}$. When the momentum $k$ sweeps through the Brillouin zone $[0, 2\pi]$, the phase accumulates $2\pi \frac{n}{d} $, corresponding to $n/d$ full counterclockwise windings. Therefore, the winding number approaches the maximal value $+n/d$. However, if $|J_3| \gg 1+ |J_2|$, $\mathbf{d}_\mu(k)$ reduces to $tJ_{3,\mu} e^{-i \frac{km}{d}}$. Here, the phase $\theta=-\frac{km}{d}$ induces a clockwise rotation, accumulating $-2\pi \frac{m}{d} $ over the period and producing $m/d$ clockwise windings. The winding number approaches the minimal value $-m/d$. Thus, for a given choice of $(n,m,\mu)$, the extremal winding numbers are 
\begin{equation}
    \omega_\mu= +\frac{n}{d} \quad \text{and} \quad  \omega_\mu= -\frac{m}{d} 
\end{equation}
realized, respectively, in the $J_{2,\mu}$-dominated and $J_{3,\mu}$-dominated limits.
\subsection{Supplementary Note 2: Semiconducting SWNTs with $n = m + 1$}\label{Note2}
To broaden the scope of the multiple phase transitions induced by the disordered long-range hopping amplitudes in semiconducting SWNTs, we now extend our analysis to the remaining semiconducting case, $n = m + 1$. As noted earlier, these nanotube systems exhibit a finite spectral gap with vanishing winding number in the absence of disorder. 

\begin{figure}[hpbt] 
$\begin{array} {cc}
\includegraphics[width=0.45\columnwidth]{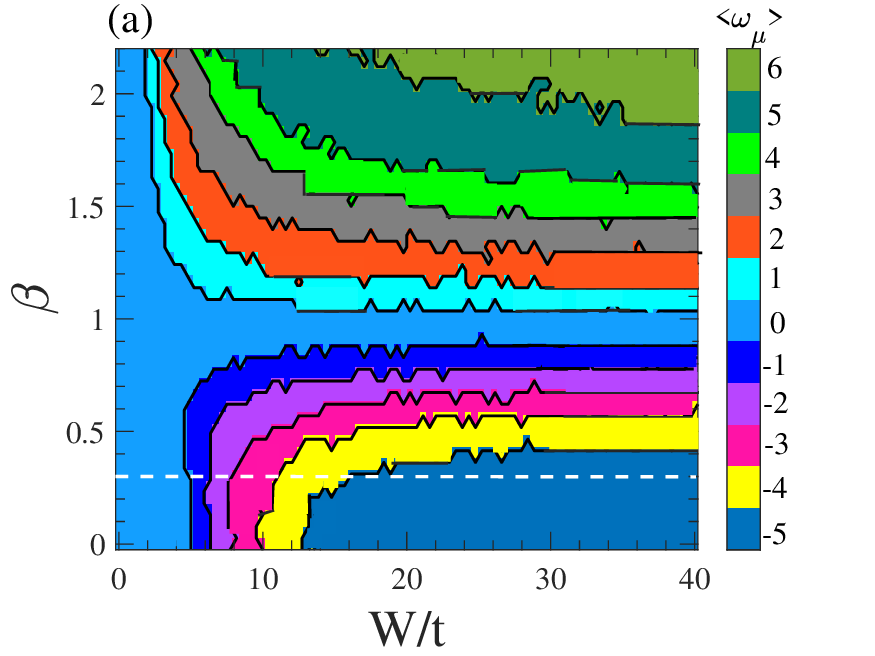}
\includegraphics[width=0.45\columnwidth]{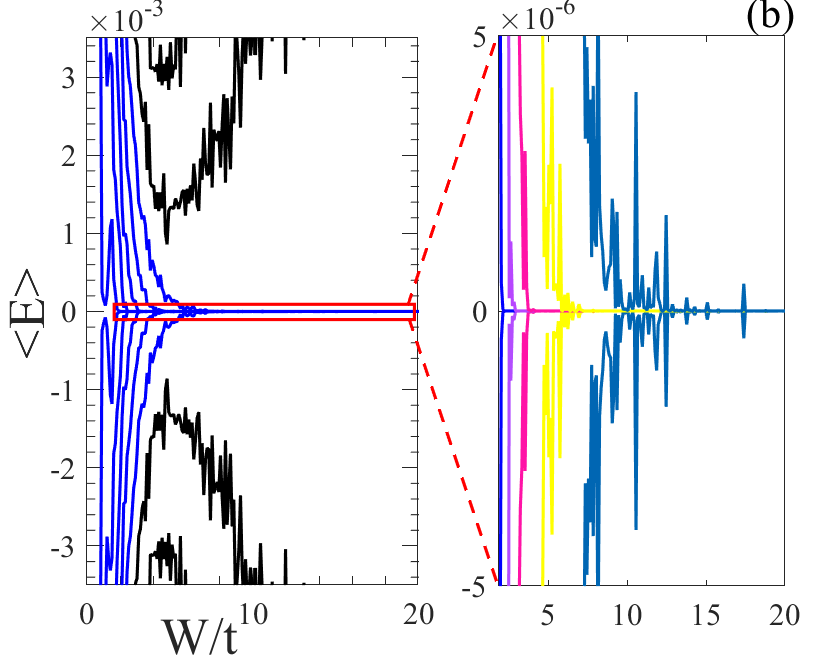}
\end{array}$
\vspace{-3mm}
\caption{(a) Disorder averaged winding number as a function of $\beta$ and $W$ for (6,5)-SWNT with $L=1000$ unit cells. $100$ samples are averaged for each point. The white dashed line corresponds to the cutting for $\beta=0.3$. (b) Averaged energy spectrum as a function of $W$ under the open boundaries conditions for $\beta=0.3$. }
\label{phaseDiag65}
\end{figure}

\smallskip 
In Fig. \ref{phaseDiag65}(a), we display the disorder-averaged topological phase diagram in the $W\text{-}\beta$ plane for $(n,m)=(6,5)$ and angular momentum $\mu=0$. With an increase of $W$, the system undergoes a one-way staircase TAI phases. For $\beta<1$ $(\beta>1)$, the disordered hopping $J_3$ $(J_2)$ dominates and the averaged winding number $\langle\omega_\mu\rangle$ decreases (increases) stepwise to $-m/d=-5$ ($n/d=+6$). Thus, depending on which long-range hopping is strongly disordered, the system realizes a robust TAI staircase that flows either towards a large positive or negative winding number. In contrast, for $\beta\approx1$, where disorder acts symmetrically on $J_2$ and $J_3$, no staircase is observed and $\langle\omega_\mu\rangle$ remains zero for all $W$. 

\smallskip 
Figure \ref{phaseDiag65}(b) shows the open-chain energy spectrum as a function of $W$ for $\beta=0.3$, i.e., in a $J_3$-dominated regime. Starting from a trivial insulator without midgap states at small $W$, increasing the disorder amplitude produces additional eigenvalues that pin at $E=0$, each corresponding to a new pair of edge-localized modes. The successive accumulation of zero-energy states in the spectrum mirrors the stepwise evolution of $\langle\omega_\mu\rangle$. 

\smallskip
Taken together with the results for type-1 and type-2 SWNTs, these data show that the disorder-induced TAI staircase and its saturation at $\langle\omega_\mu\rangle = +n/d$ or $-m/d$ are generic features of semiconducting SWNT-like chiral chains. By tuning $\beta$, one can realize cascades to either high positive or negative winding numbers, starting from a trivial clean phase. 

\subsection{Supplementary Note 3: Finite-size scaling}\label{Note3}
\begin{figure}[hpbt] 
$\begin{array} {cc}
\includegraphics[width=0.45\columnwidth]{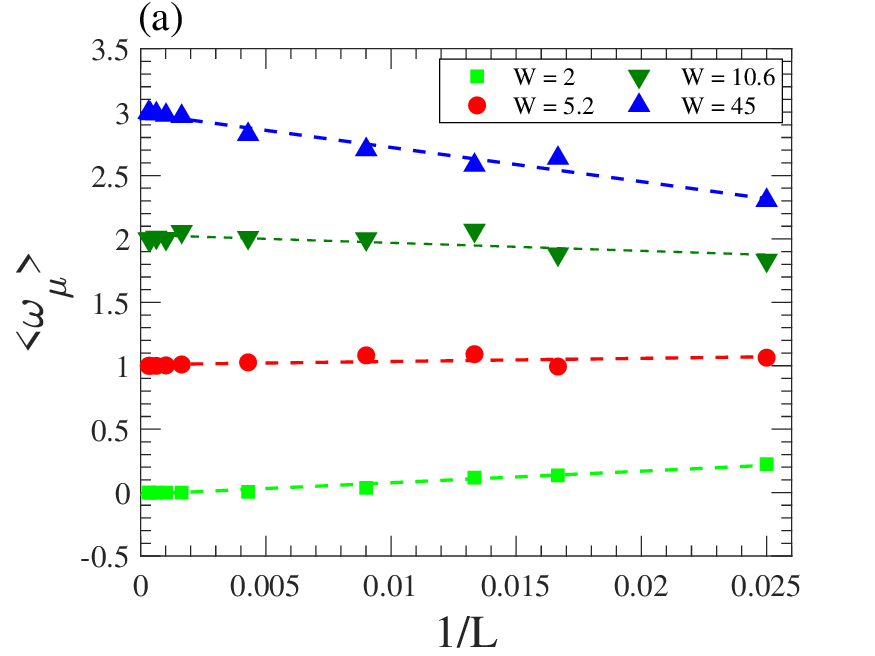}
\includegraphics[width=0.45\columnwidth]{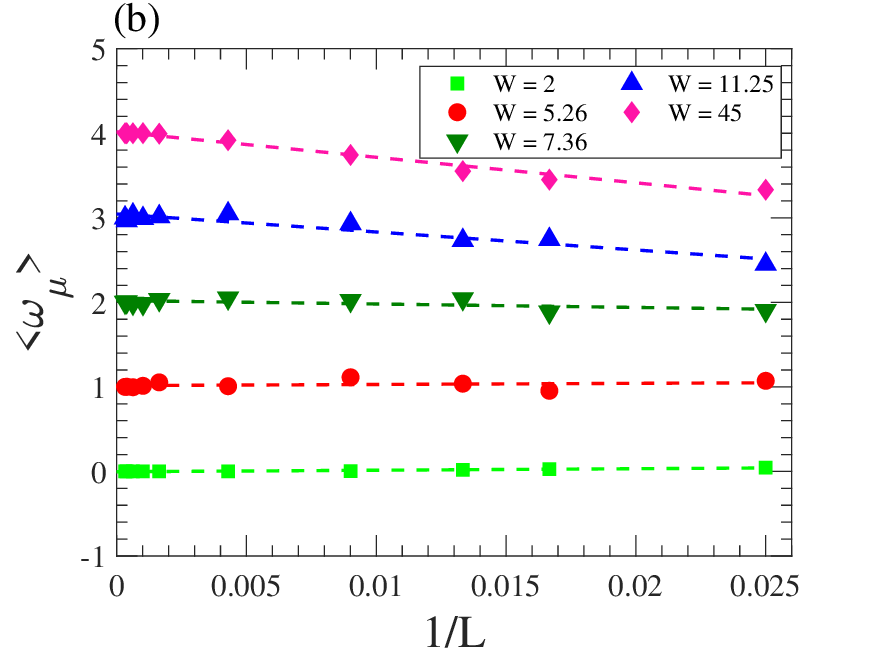}
\end{array}$
\vspace{-3mm} 
\caption{System-size $L$ dependence of the disorder-averaged winding number for different disorder strength $W$. (a) $(n,m,p,q, \beta)= (12,4,1,0, 2)$ and (b) $(n,m,p,q, \beta)= (8,6,-1,-1, 2.8)$. Dashed lines are linear fits of the numerical data.}
\label{scaling}
\end{figure}

In this section, we examine the convergence of the disorder-averaged winding number for various lattice sizes $L$. In Figs. \ref{scaling}(a) and \ref{scaling}(b), we show the variation of $\langle\omega_\mu\rangle$ as a function of $1/L$, for initially trivial topological states with $(n,m)=(12,4)$ and $(8,6)$, respectively. We have considered $10^3$ disorder realizations for each $L$.

\smallskip 
One can clearly see that for small systems $(L<100)$, the calculated winding number shows large fluctuations. This is because the edge states hybridization on opposite boundaries, resulting in an energy splitting which shifts their energies away from zero energy. As $L$ increases, this hybridization is suppressed and the zero modes become exponentially well localized at the ends. Once $L \geqslant 100$, $\langle\omega_\mu\rangle$ converges tightly to the quantized values. In the main text, all numerical results are therefore obtained for chains with $L=1000$ unit cells, which are effectively in the thermodynamic regime. Our experimental topolectrical realization is limited to $L=95$ unit cells. The finite-size scaling in Fig. \ref{scaling} shows that this length already very close to the converged regime.

\smallskip 
To further validate the convergent behavior, we consider the lowest-order finite-size scaling form, satisfying 
\begin{equation}
    \langle\omega_\mu\rangle = \frac{a}{L}+\langle\omega_\mu^\infty\rangle\,,
\end{equation}
where the slope $a$ provides an optimal fit of the original data. Table \ref{extrapolated-winding} presents the extrapolated winding number in the thermodynamic limit $\langle\omega_\mu^\infty\rangle$ for particular values $W$ in different topological phases.  
\begin{table}[h]
    \centering
\begin{tabular}{ccccc|ccccc}
\hline
\bm{$(n,m, \mu)$} &\multicolumn{4}{c|}{$(12,4, 2)$} & \multicolumn{5}{c}{$(8,6, 0)$}  \\
\hline \hline
\textbf{W} & 2& 5.2 & 10.6& 45 & 2 & 5.26& 7.36 & 11.25&45\\
\hline
\bm{$\langle\omega_\mu^\infty\rangle$} & 0.014&1.009 & 2.012& 3.001  & 0.004 & 1.015& 2.022 & 3.005&4.017\\
\hline
\end{tabular}
    \caption{Extrapolated results for disorder-averaged winding number $\langle\omega_\mu^\infty\rangle$ in the presence of several disorder amplitude $W$. Other parameters are the same as in Fig. \ref{scaling}. }
    \label{extrapolated-winding}
\end{table}
\subsection{Supplementary Note 4: Self-consistent Born approximation}\label{Note4}
In this section, we provide a detailed discussion of the SCBA used to explain the averaged bulk gap closing under disorder, as shown in Fig. 1(b) in the main text.\\

Disorder is introduced into the two long-range intercell hopping amplitudes of each unit cell $l$ of the form 
\begin{equation}
    W_1\sum_{l}^{N} \epsilon_l^{(2)} U_2(l)+W_2\sum_{l}^{N} \epsilon_l^{(3)} U_3(l)\,,
\end{equation}
where the matrix elements of each long-range bond operator $U_j(l)$ are given by
\begin{equation}
    [\,U_j(l)\,]_{\alpha\beta}
= \delta_{\alpha,(A,l)}\,\delta_{\beta,(B,l+\Delta'_j)}
\;+\;
\delta_{\alpha,(B,l+\Delta'_j)}\,\delta_{\beta,(A,l)},
\quad j=2,3.
\end{equation}
Here we assume that the random variables are taken from a uniform distribution in the interval $[-1/2, 1/2]$ without correlations as 
\begin{equation}
    \big\langle \epsilon_l^{(j)} \big\rangle =0, \quad 
    \big\langle \epsilon_l^{(j)} \epsilon_{l'}^{(r)}\big\rangle = \tfrac{1}{12} \delta_{jr} \delta_{ll'}, \quad j,r=2,3,
\end{equation}
with $\langle \cdots \rangle$ denoting the ensemble average over different disorder configurations.

\begin{figure}[hpbt] 
\includegraphics[width=0.45\columnwidth]{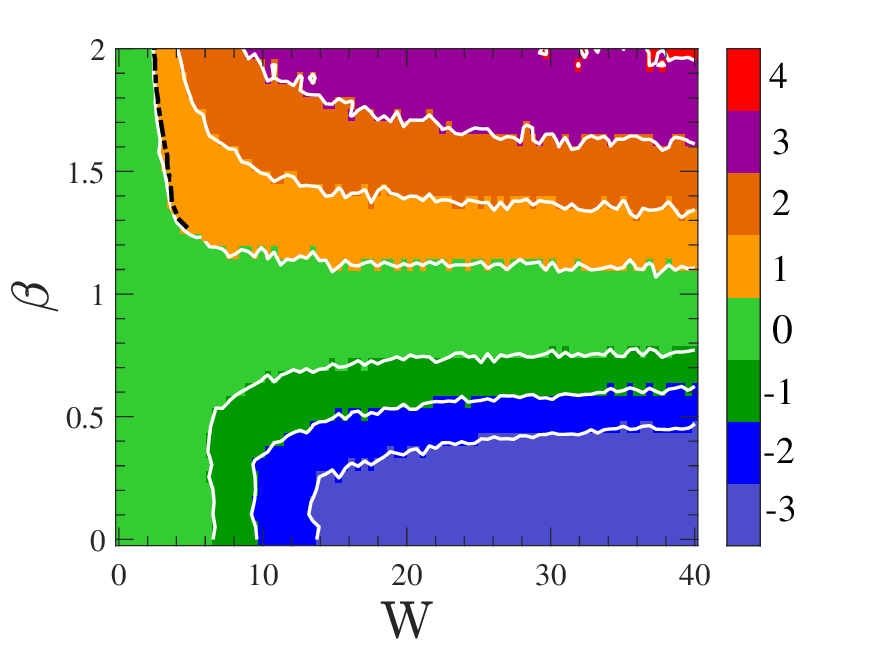}

\vspace{-3mm} 
\caption{Phase diagram of the $(8,6)$-SWNT in the W-$\beta$ plane. Every data point on the figure is averaged over 100 samples. Black dash curves are the boundaries of the first topological phase transition,  determined via the SCBA analysis.}
\label{successiveTopo}
\end{figure}

\smallskip 
For a weak disorder strength, the effective Hamiltonian at $E=0$ is given by $H_{eff} (k)= H_0 (k)+ \Sigma (E=0)$ where $H_0 (k)$ is the clean Bloch Hamiltonian \eqref{cleanH}. The self-energy $\Sigma$ in the presence of off-diagonal disorder is computed self-consistently as
\begin{equation}\label{Self}
    \Sigma (E) = \sum_{j=2,3} \tfrac{1}{12}{W_j^2} \sum_{k \in BZ} U_j G U_j\,,
\end{equation}
where $G=[(E+i 0^+) \mathcal{I} - H_0 (k) - \Sigma (E)]^{-1}$ is the dressed Green's function and \begin{equation}\label{Hreduced}
    U_2 =
\begin{pmatrix}
0 & \exp(i \frac{n}{d} k) \\
\exp(-i \frac{n}{d} k) & 0 
\end{pmatrix}, \quad U_3 =
\begin{pmatrix}
0 & \exp(-i \frac{m}{d} k) \\
\exp(i \frac{m}{d} k) & 0 
\end{pmatrix}. 
\end{equation}
Our numerical result indicates that the self-energy can be parameterized as $\Sigma (0)= \Sigma_x \sigma_x$ where $\Sigma_x$ is a real number. The two nonvanishing components effectively renormalize the intercell hoppings as 
\begin{equation}
    \tilde{J_2}\rightarrow t \exp^{i 2 \pi \mu \Delta^{\prime \prime}_2/d} + \Sigma_2 \,, \quad \tilde{J_3}\rightarrow t \exp^{i 2 \pi \mu \Delta^{\prime \prime}_3/d} + \Sigma_3\,.
\end{equation}
Finally, we arrive at the effective Bloch Hamiltonian as
\begin{equation}\label{Heffec}
    H_{\text{eff},\mu}(k) = t 
\begin{bmatrix}
0 & \tilde{f}_\mu(k) \\
\tilde{f}^\dagger_\mu(k) & 0 
\end{bmatrix},
\end{equation}
where $\tilde{f}_\mu(k) = t+ \tilde{J_2}\exp(ikn/d)+ \tilde{J_3}\exp(-ikm/d)$. The bulk energy gap closing that determines the renormalized topological phase boundary is then given by
\begin{eqnarray}
    1+(J_{2,\mu}+\Sigma_2) \cos\Big(\frac{k_0n}{d}\Big)+ (J_{3,\mu}+\Sigma_3) \cos\Big(\frac{k_0m}{d}\Big)&=& 0\nonumber\\  
    (J_{2,\mu}+\Sigma_2) \sin\Big(\frac{k_0n}{d}\Big)- (J_{3,\mu}+\Sigma_3) \sin\Big(\frac{k_0m}{d}\Big)&=& 0\,.
\end{eqnarray}
Hence, the critical self-energies are
\begin{eqnarray}
    \Sigma_{2,c}(k_0) &=& -\frac{\sin\big(\frac{m}{d}k_0\big)}{\sin\big(\frac{n+m}{d}k_0\big)}-J_{2,\mu}\nonumber\\  
    \Sigma_{3,c}(k_0) &=& -\frac{\sin\big(\frac{n}{d}k_0\big)}{\sin\big(\frac{n+m}{d}k_0\big)}-J_{3,\mu}\,.
\end{eqnarray}
For each $\beta$, we numerically solve Eq. \eqref{Self} for $k \in [0, 2\pi]$ to determine $\Sigma_{2}$ and $\Sigma_{3}$. By increasing disorder strength $W$, the critical value $W_c$ is determined by $\Sigma_{j}(W_c, \beta)=\Sigma_{j,c}(k_0)$ where $j=2,3$. In Fig. \ref{successiveTopo}, the results based on the SCBA for weak disorder $(W \leqslant 5t)$ are consistent with those obtained via disorder-averaged winding number. This indicates that the SCBA can explain the first topological phase transition accompanied by the bulk gap closing. The other successive transitions that emerge in the stronger disorder regime cannot be captured within the disorder-induced self-energy.
\subsection{Supplementary Note 5: Arithmetic and geometric mean DOS}\label{Note5}
To verify that the zero-energy edge states in the various TAI phases remain protected by a mobility gap, we evaluate the arithmetic mean DOS $\rho_{\rm ave}$ and the geometric mean DOS $\rho_{\rm typ}$ at $E=0$.
These are given by \cite{arithmeticDOS1,arithmeticDOS2}
\begin{equation}
    \rho_{\rm ave}(E)= \langle\!\langle \, \rho(i,E)\, \rangle\!\rangle \,, \quad 
    \rho_{\rm typ}(E)= \exp \big[\langle\!\langle \, \ln{\rho(i,E)} \, \rangle\!\rangle \big],
\end{equation}
where $\langle\!\langle \,\cdots\, \rangle\!\rangle$ stands for the arithmetic averaging over the sample sites and disorder configurations. The LDOS at site $i$ for a given energy $E$ is $\rho(i,E)=\sum_{j=1}^L |\braket{i|j}|^2 \delta(E-E_j)$ where in a finite-size system, the Dirac function is approximately replaced by $\delta(E-E_j)=\frac{\eta}{\pi[(E-E_j)^2+\eta^2]}$ with $\eta$ being a small artificial broadening parameter. Here, we use an exact diagonalization of the lattice Hamiltonian under periodic boundary conditions. 

\smallskip 
Generally, both the arithmetic and geometric averages of the LDOS vanish when the system exhibits an ordinary bulk gap \cite{arithmeticDOS1,arithmeticDOS2}. In contrast, in a TAI the gap is filled by disorder-induced bulk localized states, so $\rho_{\rm ave}$ tends to be finite, while $\rho_{\rm typ}$ remains zero. This mismatch is the hallmark of the emergence of the mobility gap, replacing the bulk-spectral gap of the clean limit \cite{UngappedTAI}.

\begin{figure}[hpbt] 
$\begin{array} {cc}
\includegraphics[width=0.45\columnwidth]{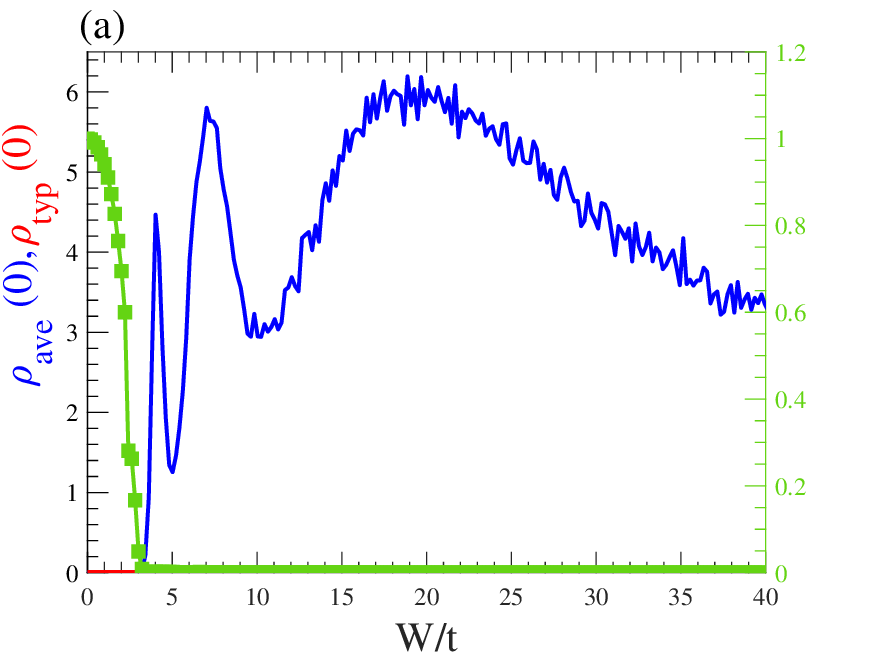}
\includegraphics[width=0.45\columnwidth]{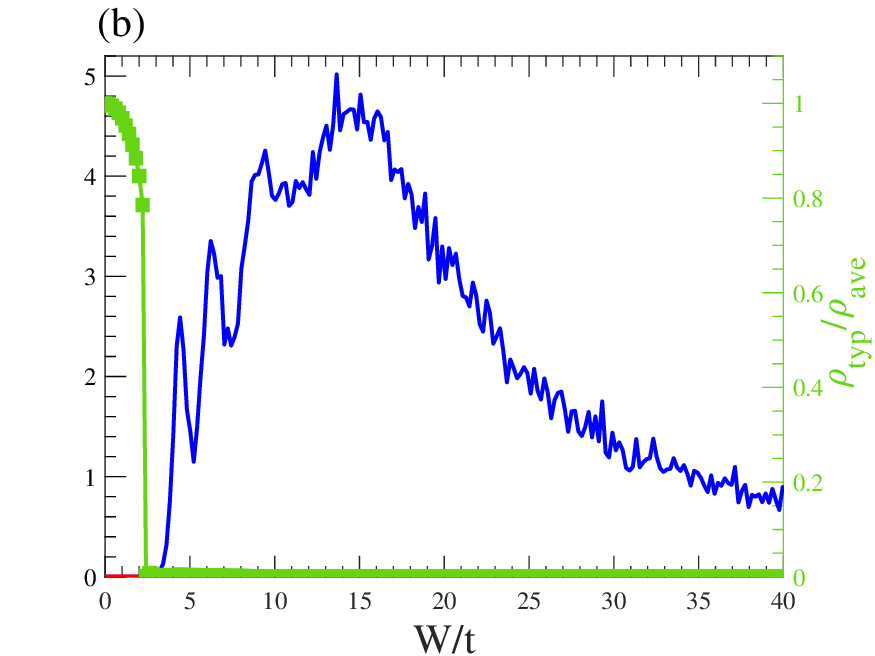}
\end{array}$
\vspace{-3mm} 
\caption{Arithmetic DOS $\rho_{ave}$ and  geometric DOS $\rho_{type}$ at zero energy versus the disorder strength $W$ for (a) $(n,m,p,q, \beta)= (12,4,1,0, 2)$ and (b) $(n,m,p,q, \beta)= (8,6,-1,-1, 2.8)$. The results are computed with the exact diagonalization method for a system size $L=4000$ and are averaged over $100$ random disorder realizations. }
\label{geomDOSE0}
\end{figure}

\smallskip 
Figures \ref{geomDOSE0}(a) and \ref{geomDOSE0}(b) display the behavior of $\rho_{\rm ave}$, $\rho_{\rm typ}$ and the ratio $\rho_{\rm typ}/\rho_{\rm ave}$ as a function of $W$ for the $(12,4)$ and $(8,6)$ SWNTs, respectively. In the weakly disordered normal insulator, $\rho_{\rm ave}\approx 0$ in the vicinity of $E=0$. Note that $ \rho_{\rm typ}/\rho_{\rm ave}$ is close to $1$ due to the small numerical broadening $\eta=10^{-4}$, which induces a uniformly distributed LDOS across the gap. When $W>W_c^{(12,4)}$, the bulk gap closes and the system undergoes a first topological phase transition into a TAI phase. At this point, $\rho_{\rm ave} (0)$ jumps to a finite value and $ \rho_{\rm typ}/\rho_{\rm ave}$ is strongly suppressed, indicating the opening of a mobility gap. Each further disorder-induced topological phase transition from a TAI with winding number $\langle\omega_\mu\rangle$ to $\langle\omega_\mu\rangle +1$ is well captured by a sharp peak in $\rho_{\rm ave} (0)$, which is consistent with that revealed by the localization length shown in Fig. 4 of the main text. Once the last winding number is reached $\langle\omega_\mu\rangle=+n/d$, $\rho_{\rm ave} (0)$ shows no additional peak and $\rho_{\rm typ} (0)$ remains vanishing for a stronger disorder, indicating the persistence of the mobility gap that protects zero-energy edge modes.
\subsection{Supplementary Note 6: Intracell-disorder case }\label{Note6}
In the main text, we focus on the case where disorder acts only on the long-range intercell hoppings $J_{2,\mu}$ and $J_{3,\mu}$, while the intarcell bond $J_1$ is kept clean. There we find a robust staircase of topological Anderson phases with disorederd- averaged winding number $\langle\omega_\mu\rangle$ increasing in integer steps and saturating at a high-index plateau.

\begin{figure}[hpbt] 
\includegraphics[width=0.45\columnwidth]{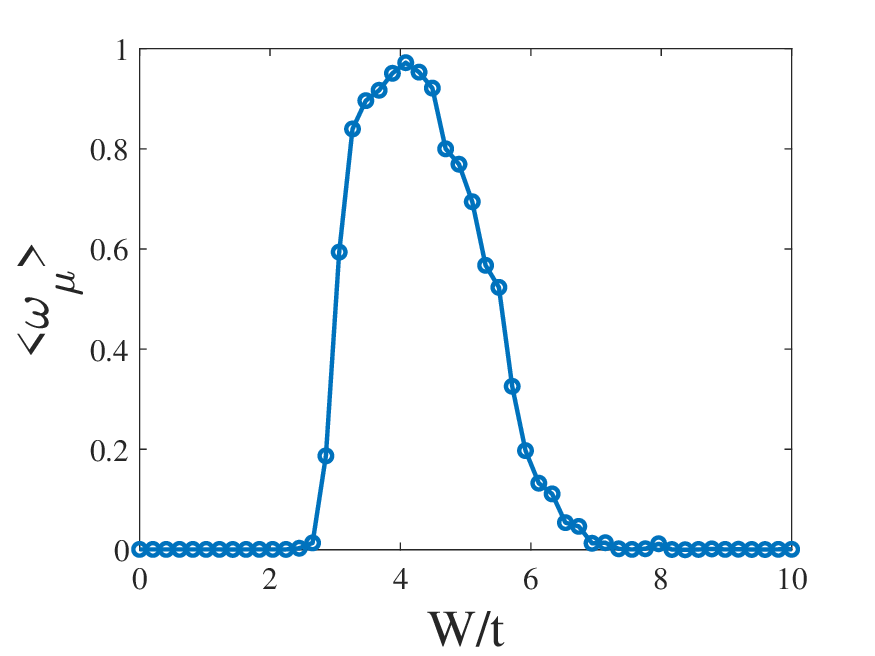}
    \caption{Disorder-averaged winding number $\langle\omega_\mu\rangle$ versus intracell disorder strength $W$ for the $(8,6)$- SWNTS. }
    \label{disordredIntra}
\end{figure}

Here, we consider the opposite situation where disorder is applied only to the intracell hopping while the long-range couplings remain clean. Concretely, for the type-2 $(8,6)$ semiconducting nanotube and the chain with angular-momentum $\mu=0$, we take
\begin{equation}
    J_{1}=t + W \epsilon, \quad J_2=t, \quad J_3=t
\end{equation}
where $\epsilon$ is random variable uniformly distributed in $[-1/2, 1/2]$. The chiral symmetry is preserved, so the  real-space winding number is still well defined and computed as in the main text. Figure \ref{disordredIntra} shows the disorder-averaged winding number 
$\langle\omega_\mu\rangle$ as a function of the intracell disorder strength $W$. 
For small $W$, the system remains a trivial insulator with $\langle\omega_\mu\rangle =0$. As $W$ is increased, a single topological transition occurs at a critical value $W^{(1)}_c$, where $\langle\omega_\mu\rangle$ rises to $1$, signaling the emergence of a 1D TAI phase. Upon further increase of diosrder amplitude, a second critical scale $W^{(2)}_c$ is reached, beyond which $\langle\omega_\mu\rangle$ decreases back toward $0$. For $W > W^{(2)}_c$, the chain behaves as a trivial Anderson insulator with all states localized and no protected edge modes.
\subsection{Supplementary Note 7: Persistence of higher winding number topological Anderson phases}\label{Note7}
As pinpointed in the main text, we now support our conjecture about the robustness of the TAI phases with higher winding number analytically. For this purpose, we consider the limit case where the disordered intercell hopping $|J_2|$ dominates, corresponding to $\beta \gg 1$. 

\smallskip 
For large enough $|J_2|\gg |J_3|$, the zero-modes Schr\"{o}dinger equation, $H\ket{\psi} = 0$, reduces to the one-step recursions as 
\begin{equation} \label{recursions}
    \phi_{j+\Delta_2'}^A=-\frac{t}{J_{2,j}} \phi_{j}^A,\qquad
\phi^{B}_{j+\Delta_2'}=-\frac{J_{2,j}}{t}\,\phi^{B}_{j}.
\end{equation}
Because this relation connects unit cell indices that are equivalents (mod $\Delta_2')$, we can rewrite the index  $j=k \Delta_2'+i_0$ where $i_0=0, 1, \cdots,\Delta_2'-1$. For a given $i_0$, the corresponding probability distributions of the zero-mode wave functions read 
\begin{equation}
    \phi_{i_0+L\Delta_2'}^A=(-1)^{L} \prod _{k=0}^{L-1}\frac{t}{J_{2,i_0+k\Delta_2'}} \phi_{i_0}^A, \qquad
\phi^{B}_{i_0+L\Delta_2'}
= (-1)^L\prod_{k=0}^{L-1}\frac{J_{2,i_0+k\Delta_2'}}{t}\phi^{B}_{i_0}\,.
\end{equation}
Hence, one can obtain the Lyapunov exponent associated with the zero energy mode for $L\rightarrow \infty$ as 
\begin{equation}
\gamma  = -\lim_{L\to\infty}\frac{1}{L}\ln\bigg|\frac{\phi_{i_0+L\Delta_2'}}{\phi_{i_0}}\bigg|\,.
\end{equation}
Generally, $\gamma$ is positive for spatially localized edge modes and tends to zero at the topological
transition point. According to Birkhoff’s ergodic theorem for uniform disorder distribution, we can use the ensemble average to evaluate $\gamma$ as
\begin{eqnarray}\label{conditions}
    \gamma_A&= &\lim_{L\to\infty} \frac{1}{L} \sum_{k=0}^{L-1} 
    \big(\ln|J_{2,i_0+k\Delta_2'}|-\ln|t|\big)
    \nonumber\\
&= & \big\langle \ln |J_{2,i_0+k\Delta_2'}|  \big\rangle_{i_0}  -\ln|t|
\nonumber\\
&= & - \gamma_B\,.
\end{eqnarray} 
Here the average $\langle \, \cdot \, \rangle_{i_0} $ is independent of $i_0$ and is given by 
\begin{equation} \label{integral}
    f(W) :=\big\langle \ln |J_{2,k}| \big\rangle = \frac{1}{2}\int_{-\frac{1}{2}}^{\frac{1}{2}} \ln \big[(t \cos{x}+ W \epsilon)^2+ (t \sin{x})^2\big] d\epsilon \,,
\end{equation}
with $x=\frac{2\pi \mu p}{d}$. Note that this function can be approximated as
\begin{equation}
f(W) \approx \left\{ 
    \begin{array}{ll}
        \ln t - \frac{\cos{2x}}{24t^2} W^2 + \mathcal{O}(W^4), & W\rightarrow 0\,, \\
        \ln W-(1+\ln 2) + \mathcal{O}(1), & W\rightarrow \infty\,.
    \end{array} 
\right.
\end{equation}
Therefore, $f(W)$ either increases monotonically from $\ln t$ when $\cos{2x} \leq 0$, or has a single global minimum and then increases for large $W$ where $\cos{2x} > 0$. Moreover, for all sufficiently large $W$, the condition $f(W)> \ln |t|$ is always satisfied, so $\gamma_A >0$ and the corresponding zero mode remains localized. Following Eq. \eqref{recursions}, the recursion decouples the system into $\Delta_2'$ independent residue classes. Each channel that satisfies $\gamma_A >0$ contributes one exponentially localized zero mode per edge, and the TAI with the winding number $ \overline{\omega}= \Delta_2' = +\frac{n}{d}$ persists for strong disorder. The same argument holds when $|J_3|$ dominates.

\bibliography{biblio.bib}
\end{document}